\documentclass[aps,twocolumn,groupedaddress]{revtex4}
\usepackage{graphicx}

\begin{document}

\title{Elementary Excitations of Ferromagnetic Metal Nanoparticles}


\author{A.~Cehovin}
\affiliation{Division of Solid State Theory, Department of Physics,
Lund University, SE-223 62 Lund, Sweden}
\author{C.M.~Canali}
\affiliation{Department of Technology, Kalmar University, 391 82 Kalmar}
\author{A.H.~MacDonald}
\affiliation{Department of Physics, University of Texas at Austin,
Austin TX 78712}


\date{\today}

\begin{abstract}

We present a theory of the elementary spin excitations in
transition metal ferromagnet nanoparticles which achieves a unified
and consistent quantum description of both collective and quasiparticle physics.
The theory starts by recognizing the essential role played by spin-orbit interactions
in determining the energies of ferromagnetic
resonances in the collective excitation spectrum and the strength of their
coupling to low-energy particle-hole excitations.
We argue that a crossover between Landau-damped ferromagnetic resonance and pure-state
collective magnetic excitations occurs as the number of atoms in
typical transition metal ferromagnet nanoparticles drops below approximately
$10^4$, approximately where the single-particle level spacing, $\delta$,
becomes larger than , $\sqrt{\alpha} E_{\rm res}$, where $E_{\rm res}$ is the 
ferromagnetic resonance frequency and $\alpha$ is the Gilbert damping parameter. 
We illustrate our ideas by studying the properties of semi-realistic
model Hamiltonians, which we solve numerically for 
nanoparticles containing several hundred atoms.
For small nanoparticles, we find one isolated ferromagnetic
resonance collective mode below the lowest particle-hole excitation energy,
at $E_{\rm res} \approx 0.1$ meV. The spectral weight of this
pure excitation nearly exhausts the transverse dynamical susceptibility spectral
weight.  As $\delta$ approaches $\sqrt{\alpha} E_{\rm res}$,
the ferromagnetic collective excitation is more likely to couple strongly with discrete
particle-hole excitations.  In this regime the distinction between
the two types of excitations blurs.  We discuss the significance of this picture
for the interpretation of recent single-electron tunneling experiments.

\end{abstract}

\maketitle


\section{Introduction}

In bulk condensed matter systems, normal metals are Fermi liquids; 
their low-energy excitation 
spectra can be placed in one-to-one correspondence with those of corresponding
non-interacting electron systems as argued by Landau more than 50 years ago.
Recent single-electron tunneling spectroscopy studies of metallic
nanoparticles \cite{delft_ralph01:_spect}, 
in which the discrete excitation spectra 
of small systems containing less than one thousand 
to tens of thousands of atoms are investigated,
have allowed Landau's enormous simplification of interacting fermion physics
to be examined quite directly.  With a few caveats and some exceptions, the 
wide variety of interesting phenomena that have been studied using single-electron
tunneling spectroscopy can be understood using ideas from independent particle
quantum mechanics.  Although they can still be regarded as Fermi liquids for 
many purposes, metals with nearly continuous broken symmetries (in particular the 
ferromagnetic transition metals that are the focus of this article) support low-energy 
collective excitations in addition to Landau's particle-hole excitations.  When
spin-orbit coupling is neglected, the broken symmetry of itinerant electron ferromagnets
is continuous and the collective excitations are Goldstone bosons whose energy 
vanishes in the long-wavelength limit.  Recent single-electron tunneling spectroscopy 
studies\cite{gueron1999,deshmukh2001} have for the first time succeeded in resolving the 
excitation spectra of ferromagnetic transition metal 
nanoparticles with diameters below 4 nm.  The 
discrete resonances seen in the tunneling experiments measure the 
low-energy many-electron excitation spectra of a 
single-domain ferromagnetic nanoparticle.  The ultimate objective of this paper is 
to shed light on the physics that is responsible for the rich and complex
behavior seen experimentally, which includes hysteretic behavior, 
non-monotonic field dependences, and a much denser low-energy excitation spectrum
than would be expected based on a non-interacting quasiparticle model.
Our approach is based on a weak-coupling
description of a metallic ferromagnet in which spin-orbit interactions cause 
collective and particle-hole excitation to be coupled at low-energies, and 
the classical micromagnetic description of a ferromagnetic metal nanoparticle
emerges naturally when quantum effects are unimportant.  

Our theory builds on earlier work \cite{cmc_ahm2000prl,kleff2001prb}, 
which captures several features of the experimental spectra, 
especially when non-equilibrium excitations\cite{kleff_vdelft2001prb}
are considered, but does not provide a unified and 
consistent quantum description of how collective and quasi-particle excitations  
are coupled by spin-orbit interactions.
The purpose of this paper is to
develop such a description and illustrate its implications by 
applying a simplified but qualitatively realistic microscopic model
that we have recently introduced\cite{ac_cmc_ahm2002}.

The elementary spin excitations in bulk itinerant-electron ferromagnets
are of two kinds: collective spin excitations (spin-waves) and
spin-flip particle-hole excitations (Stoner excitations).  
Spin-wave excitations
are related to the collective magnetization degree of freedom and 
form a branch in $q-\omega$ space, which is gapless if
the system is isotropic in accord with Goldstone's theorem. 
The main effects of spin-orbit interactions in the bulk 
are to generate an energy gap $E_{\rm res}$
in the $q=0$ collective mode, which is of the order of the consequent 
magneto-crystalline anisotropy energy per atom -- 
$K \approx 0.1$ meV in Cobalt\cite{stearns1986} --
and to introduce the possibility of decay of 
long-wavelength collective excitations into particle-hole
excitations\cite{korenman_prange1972, korenman1974}, 
a process that contributes substantially 
to the collective excitation lifetime%
\footnote{In this paper we will focus on the damping of the 
uniform collective mode arising from the decay into particle-hole
excitations. There are of course other important sources of damping
not considered here, for example magnetoelastic scattering by phonons
and magnon-magnon interactions enhanced by surface defects.
Some of these relaxation mechanisms have been
investigated recently both experimentally\cite{ingvarsson2002} 
and theoretically\cite{safonov_bertram2002}.} for the case of NiFe thin films.
In the absence of spin-orbit interactions, the ferromagnetic resonance 
is coupled only to spin-flip particle-hole excitations
which at long wavelengths have a gap of the order of 
the spin-splitting field $\Delta$. Gapless spin-flip particle-hole 
excitations are possible
only at wavevectors exceeding the minimum $q$-space separation between
majority and minority spin Fermi surfaces. 
The separation in energy at long-wavelengths between
collective modes and spin-flip particle-hole continuum
implies that low-energy collective modes are only weakly damped.
Beyond a critical value of $q$ the spin-wave 
branch merges with the continuum.
Thus spin-waves can decay into Stoner
excitations.  The strength of this decay process is sensitive 
to the character of the orbitals involved in the particle-hole
excitations\cite{cooke1980}.
In the present work we will focus on how this description of
elementary spin excitations in itinerant ferromagnets has to be altered
when the level spacing for quasi-particle excitations $\delta$ is finite 
and approaches $E_{\rm res}$,
a condition that is satisfied in the nanometer particle-size range. Note that
$\delta$ is inversely proportional to particle volume, while
$E_{\rm res}$ is approximately volume independent for large nanoparticles.
Since in a finite system there is no wavevector
conservation, collective modes and spin-flip particle-hole pairs cannot be 
simply separated, unlike the bulk case.
We will show that this fact, together with the essential role played
by spin-orbit interaction, has profound consequences on the nature of the
elementary spin excitations in ferromagnetic metal nanoparticles.

The paper is organized in the following way.  In Sec.~\ref{models}
we introduce two similar microscopic models for a magnetic metal nanoparticle
and explain how they are related to the phenomenological
model consider previously.  The models differ in that one accounts for 
the difference in the strength of exchange interactions between s and 
d electrons in transition metals, a feature whose consequences we wish to
address specifically.  In Sec.~\ref{path_integral} we derive  
a path integral formulation of theories based on
these models\cite{ahm_cmc2001ssc}.  This point of view provides a convenient 
language for explaining the interplay between collective modes and particle-hole
excitations, and for making contact with classical micromagnetic
theory. The spin-orientation fluctuation propagator
in the Gaussian approximation 
is discussed in Sec.~\ref{sec_gauss}. The poles of the fluctuation propagator  
occur at the elementary spin excitations of the system.
We will show that the ferromagnetic resonance energy $E_{\rm res}$ 
can be expressed as the quotient of anisotropy
energy and Berry curvature coefficients 
which specify the Gaussian expansion of the action at low frequencies.  
 \begin{figure}
 \includegraphics[width=2.4in,height=2.9in]{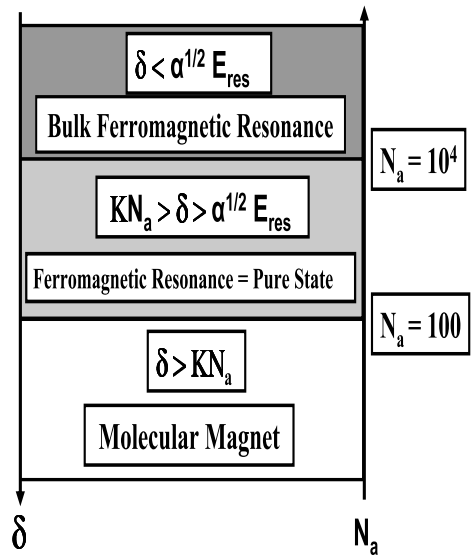}
 \caption{Crossing of relevant energy scales as a function of the
number of atoms ${\cal N}_a$ in a magnetic nanoparticle. Here
$\delta$ is the single-particle mean level spacing; $E_{\rm res}$ is
the energy of the coherent spin collective mode or
ferromagnetic resonance energy; $K$ is the magneto-crystalline anisotropy
energy per atom, and $\alpha$ is the Gilbert damping factor.}
 \label{fig:one}
 \end{figure}
We then discuss how the resonance evolves with particle-size.  Our 
main conclusions is summarized in Fig.~\ref{fig:one}.  For large particle-sizes
the ferromagnetic resonance weight is distributed over a large number of 
particle-hole excitations, while for smaller particle sizes the ferromagnetic
resonance appears as a pure quantum excitation.  The crossover between the 
two regimes occurs 
approximately where the level spacing $\delta$ is equal to 
$\sqrt{\alpha}\; E_{\rm res}$ where $\alpha$ is the bulk resonance's 
Gilbert damping factor\cite{gilbert1955}.  
For typical transition metal nanoparticles this
condition is satisfied in particles containing of order $10^4$ atoms. 
For smaller particles, avoided crossings between collective and 
individual particle-hole excitations will occur occasionally as 
a function of system parameters, for example as a function of an external
magnetic field used to reorient the magnetization.  For nanoparticles 
containing fewer than of order $10^2$ atoms, avoided crossings with 
particle-hole excitations will usually not occur at any field orientation,
and the nanoparticle can be considered to be a molecular magnet
in which only spin-orientation degrees of freedom are important at low 
energies\cite{ccmPRL03}.  
In Sec.~\ref{numerics} we will present numerical results 
for a few-hundred atom nanoparticle, which
illustrate some of the these points.
Finally in Sec.~\ref{final} we summarize our findings 
and comment on their relevance in understanding current 
tunneling experiments.

\section{Quantum models of a ferromagnetic metal nanoparticle}
\label{models}

In this article we consider two slightly different quantum models chosen to describe
both collective and quasi-particle physics in a magnetic metal nanoparticle.
We will denote them as Local D-orbital Exchange model (LDE model) and
Long-Range Exchange model (LRE model)
respectively, for reasons that will become clear below.
We will see that these models, when solved within a mean-field
approximation, are essentially equivalent
and provide a convenient quantum
description of a ferromagnetic nanoparticle when the magnetization
is {\it coherent} (spatially constant across the sample).  Our use of 
these models is motivated partly by the evident success of spin-density-functional
theory in describing ferromagnetism in transition metals; our formalism
could easily be adapted to be compatible with this method of calculating the  
energies of different magnetic configurations.  The models are 
intended to be sufficiently realistic to capture generic aspects of transition
metal nanoparticle magnetism, but evidently miss features that can be 
important in practice such as variation in exchange interaction strength 
and inter-atom hopping amplitudes near the surface of the nanoparticle.

\subsection{Local D-orbital Exchange Model}
The first model that we consider
accounts qualitatively for the orbital dependence of exchange interaction
strengths in a transition metal itinerant-electron ferromagnets \cite{ac_cmc_ahm2002}. 
We model the nanoparticle as a cluster of ${\cal N}_a$ atoms located
on the sites of a truncated crystal.  The numerical results we present here
are for a cobalt cluster whose truncated f.c.c. crystal is
circumscribed by a hemisphere whose equator lies in the $XY$-plane of the
f.c.c. crystal\footnote{\label{fnote0} We will use $X$, $Y$ and $Z$ to
indicate the directions of the f.c.c. crystal axes.}.
The choice of a hemisphere is motivated by the tunneling experiments
of Ref.~\onlinecite{gueron1999,deshmukh2001}.
We use a s-p-d tight-binding model for the quasiparticle orbitals,
with 18 orbitals per atom, including the spin-degree of 
freedom.  Nine orbitals per Co atom are occupied in neutral nanoparticles.
The full second-quantized Hamiltonian is,
\begin{equation}
\hat {\cal H}={\hat H}_{\rm band}+{\hat H}_{\rm exch}+{\hat H}_{\rm so}+
{\hat H}_{\rm Zee}\; .
\label{totham}
\end{equation}
The second-quantized one-body term

\begin{equation}
{\hat H}_{\rm band} = \sum_{i,j} \sum_{s} \sum_{\mu_1, \mu_2}
t^{i,j}_{\mu_1,\mu_2,s}
c^{\dagger}_{i,\mu_1,s}c^{\phantom{\dagger}}_{j,\mu_2,s}
\label{band}
\end{equation}

is written in terms of creation and annihilation operators 
$c^{\dagger}_{i,\mu_1,s}$ and $c^{\phantom{\dagger}}_{j,\mu_2,s}$
labeled by atomic-site indexes
$i,j$, atomic angular momentum indexes $\mu_1,\mu_2$ and spin indexes
$s,s'$. We choose the spin-quantization axis to be along the direction of 
the magnetization, which is specified by a  unit vector 
$\hat \Omega(\Theta, \Phi)$ where $\Theta$ and
$\Phi$ are the usual angular coordinates defined with 
respect to the f.c.c. crystal
axes.  The parameters $t^{i,j}_{\mu_1,\mu_2,s}$ are
Slater-Koster parameters\cite{slater_koster} obtained after performing a
L\"owdin symmetric orthogonalization procedure\cite{mattheiss}
on the set of Slater-Koster parameters for non-orthogonal atomic orbitals
of bulk spin-unpolarized Co\cite{papaconstantopoulos}.
Here the exchange term is a short-range spin-interaction
involving only the electrons spins of $d$-orbitals on the same atomic site:
\begin{equation}
{\hat H}_{\rm exch}=
-2U_{dd}\sum_{i} \vec{S}_{d,i} \cdot \vec{S}_{d,i}\; ,
\label{exchH}
\end{equation}
where
\begin{equation}
\vec{S}_{d,i} \equiv \sum_{\mu\in d } \vec S_{i,\mu}
= \sum_{\mu\in d }
\frac{1}{2}
\sum_{s,s'}c^{\dagger}_{i,\mu,s}{\vec\sigma}^{\phantom{\dagger}}_{s,s'}
c^{\phantom{\dagger}}_{i,\mu,s'}\, ,
\label{exchange}
\end{equation}
${\vec \sigma}$ being
a vector whose components
$\sigma^{\alpha}$, $\alpha=x,y,z$ are the three Pauli matrices.
The parameter $U_{dd}$ in Eq.~(\ref{exchH}) determines
the strength of the exchange interaction
and we will set it equal to $1$ eV in our numerical 
calculations\cite{ac_cmc_ahm2002}.
This choice leads to the correct magnetization per atom in the bulk.

The spin-orbit coupling ${\hat H}_{\rm so}$ is a {\it local} one-body operator
\begin{equation}
{\hat H}_{\rm so} = \xi_d\sum_i \sum_{\mu,\mu',s,s'}
\langle \mu,s| {\vec L}\cdot {\vec S}|\mu',s'\rangle
c^{\dagger}_{i,\mu,s}c^{\phantom{\dagger}}_{i,\mu',s'}\;,
\end{equation}
where the atomic matrix elements
$\langle \mu,s| {\vec L}\cdot {\vec S}|\mu',s'\rangle \equiv
\langle i, \mu,s| {\vec L}\cdot {\vec S}|i, \mu',s'\rangle$ depend 
on the spin-quantization axis specified by the angles $\Theta$ and $\Phi$.
The energy scale $\xi_d$, which characterizes the coupling between
spin and orbital degrees of freedom,  varies in the range from 50
to 100 meV in bulk 3d transition metal ferromagnets\cite{soinmetals}.
Finally, ${\hat H}_{\rm Zee}$ is a local
one-body operator describing the Zeeman coupling
of the orbital and spin degrees of freedom to an external
magnetic field $\vec{H}_{\rm ext}$:
\begin{widetext}
\begin{eqnarray}
{\hat H}_{\rm Zee}= &&-\mu_{\rm B}\sum_i \sum_{\mu,\mu',s,s'}
\langle \mu,s|(\vec{L}+g_{s}\vec{S}|\mu',s')\rangle
\cdot\vec{H}_{\rm ext}\,
c^{\dagger}_{i,\mu,s}c^{\phantom{\dagger}}_{i,\mu',s'}\nonumber\\
 =&& -\mu_{\rm B}\sum_i \vec{H}_{\rm ext}\cdot \Big\{\sum_{\mu,\mu',s}
\langle \mu,s|\vec{L}|\mu',s)\rangle
\, c^{\dagger}_{i,\mu,s}c^{\phantom{\dagger}}_{i,\mu',s}
+ {g_s\over 2}
\sum_{\mu,s,s'}c^{\dagger}_{i,\mu,s}{\vec\sigma}^{\phantom{\dagger}}_{s,s'}
c^{\phantom{\dagger}}_{i,\mu,s'}\Big\}\; ,
\end{eqnarray}
\end{widetext}
with  ${g_s = 2}$. ${\hat H}_{\rm Zee}$ plays an important role
in manipulating many-body states.
In Ref.~(\onlinecite{ac_cmc_ahm2002}) we have investigated the
spectrum of the microscopic Hamiltonian of Eq.~(\ref{totham}), treating
the quartic exchange interaction in the mean-field approximation.
We have focused in particular on the mesoscopic physics of the
quasiparticle energies and their complex behavior as a function
of the magnetization orientation and external magnetic field orientation and strength.
This analysis, although very relevant to the understanding of
tunneling experiments, does not tell us anything about the
quantization of the collective magnetization orientation dynamics.
For this purpose and for
the purpose of making a connection with the classical micromagnetic theory,
the path integral approach described in the next section provides a more useful language.

\subsection{Long-Range Exchange Model}
Our microscopic LDE model, when solved in the mean-field
approximation, is related to a toy model, originally introduced in 
Ref.~(\onlinecite{cmc_ahm2000prl}), which can be regarded as 
the simplest possible model of a ferromagnetic
metal nanoparticle.  The toy model  Hamiltonian 
assumes identical exchange constants between all pairs of single-particle
orbitals in the nanoparticle: 
\begin{widetext}
\begin{equation}
\hat {\cal H} 
= \sum_{n,s} c^{\dagger}_{n,s} c^{\phantom{\dagger}}_{n,s} \epsilon_n
- \frac{U}{8{\cal N}_a} \sum_{n,m} \sum_{s,s',t,t'} c^{\dagger}_{n,s'}
{\vec \sigma}_{s',s} c^{\phantom{\dagger}}_{n,s} \cdot 
c^{\dagger}_{m,t'} {\vec \sigma}_{t',t}
c^{\phantom{\dagger}}_{m,t}
= {\hat H}_{\rm band} - \frac{1}{2} \frac{U}{{\cal N}_a} 
\vec S\cdot \vec S\;.
\label{toyhamiltonian}
\end{equation}
\end{widetext}
In Eq.~\ref{toyhamiltonian}
$c^{\dagger}_{n,s}$ and $c^{\phantom{\dagger}}_{n,s}$ are Fermion creation
and annihilation operators for a quasi-particle state characterized by
orbital energy $\epsilon_n$ and spin component $s$;
$\vec S =  \frac{1}{2}\sum_n  c^{\dagger}_{n,s'} {\vec \sigma}_{s',s} 
c^{\phantom {\dagger}}_{n,s}$ is
the total spin of the nanoparticle.
The single
particle orbitals will have an average spacing inversely
proportional to the volume of the nanoparticle (or the number of
atoms ${\cal N}_a$) and are expected to
exhibit spectral rigidity\cite{efetov_susy}.
The one-body term
in this Hamiltonian, ${\hat H}_{\rm band}$, should be thought of as including
a mean-field approximation to those spin-independent 
portions of the interaction not
captured by the exchange term 
${\hat H}_{\rm exch}= - \frac{1}{2} \frac{U}{{\cal N}_a} \vec S\cdot \vec S$.
The many-particle spectrum of this Hamiltonian has been discussed
in detail in Ref.~(\onlinecite{cmc_ahm2000prl}).

This toy model Hamiltonian can be further augmented\cite{ahm_cmc2001ssc}
by a one-body spin-orbit coupling term $H_{\rm so}$. 
We can write ${\hat H}_{\rm so}$ as\cite{delft_ralph01:_spect}
\begin{equation}
{\hat H}_{\rm so} = \sum_{n,m,s}v^{s}_{n,m} 
c^{\dagger}_{n,s} c^{\phantom{\dagger}}_{n,{\bar s}}\;,
\end{equation}
with
\begin{equation}
v^{s}_{n,m} = \big(v^{\bar s}_{n,m})^{\star}= -v^{s}_{m,n}\;,\ \ \
v^{s}_{n,n} = 0\; ,
\label{eq:spinorbit}
\end{equation}
where the conditions on the matrix elements $v^{s}_{n,m}$ specified in
Eq.~(\ref{eq:spinorbit}) 
ensure that ${\hat H}_{\rm so}$ is hermitian and invariant
under time-reversal.

Consider the orbital part of the microscopic Hamiltonian 
given in Eq.~(\ref{totham}). 
Since ${\hat H}_{\rm band}$ is quadratic, it can be diagonalized
by a canonical transformation:
\begin{equation}
{\hat H}_{\rm band} = \sum_{i,j} \sum_{s} \sum_{\mu_1, \mu_2}
t^{i,j}_{\mu_1,\mu_2,s}
c^{\dagger}_{i,\mu_1,s}c^{\phantom{\dagger}}_{j,\mu_2,s} =
\sum_{n,s} c^{\dagger}_{n,s} c^{\phantom{\dagger}}_{n,s} \epsilon_n\;,
\label{orbit_diag}
\end{equation}
where
\begin{equation}
c^{\phantom{\dagger}}_{n,s} = \sum_{i,\mu,t} \langle{n,s}|{i,\mu,t}\rangle
c^{\phantom{\dagger}}_{i,\mu,t}\;,
\label{can_tr}
\end{equation}
and $|n,s\rangle$ are the orthonormal eigenvectors. 
The eigenvalues $\epsilon_n$ in Eq.~(\ref{orbit_diag})
can be identified
with the doubly degenerate orbital energies of the toy model Hamiltonian 
in Eq.~(\ref{toyhamiltonian}). If we now use Eq.~(\ref{can_tr}) and
the completeness of the eigenstates $\{|i,\mu,t\rangle\}_{i,\mu,t}$, we can
rewrite the exchange interaction of Eq.~(\ref{toyhamiltonian}) in the form
\begin{equation}
{\hat H}_{\rm exch}= - \frac{1}{2} \frac{U}{{\cal N}_a} \vec S\cdot \vec S
= - \frac{U}{2{\cal N}_a} \frac{1}{4}\sum_{i,j}\vec S_i\cdot \vec S_j
\label{LR_exch}
\end{equation}
with $\vec S_i \equiv \sum_{\mu} S_{i\mu}$. From Eq.~(\ref{LR_exch}) it is
clear that the toy model Hamiltonian with equal exchange constants
is equivalent to a microscopic model with a {\it long-range exchange}
interaction, coupling the spin of each atomic orbital at each atomic site
to all others; hence the name
of Long-Range Exchange model (LRE model).
We emphasize that the electron spins of all orbitals are coupled in this model,
not just the spins of $d$-orbitals. 
From Eq.~(\ref{LR_exch}) it follows that in a constrained mean-field
approximation where the local moment $\langle \vec S_i\rangle$ 
is forced to be {\it coherent} across the sample 
and the spins of {\it all} atomic orbitals 
are exchange-coupled, the microscopic LDE model and the
the LRE model would be equivalent equivalent, 
(with $2U_{dd} \Leftrightarrow U/8$).  
Differences arise when only the
the $d$-orbitals in the LDE model are exchange-coupled,
which we comment on below.

Neither model includes any magnetostatic dipole-dipole interactions,
which can be important in some circumstances, for example when
the nanoparticle is not close to spherical, but are easily incorporated 
in our discussion.  Note however, that
both models lead to strong shape dependence,
because of surface effects which are important when the particle size is small.
In Sec.~(V) we will see that the structure
of particle-hole excitations of the LDE model when only 
the spins of d-orbitals are coupled
is richer than when quasiparticle majority and minority
spins are simply shifted by a rigid exchange field, as in the case of
the LRE model.

\section{Auxiliary Field Formulation} 
\label{path_integral}

\subsection{Coherent state functional integral and 
Hubbard-Stratonovich transformation}
We consider first our LRE toy model. The extension to the LDE model is
straightforward and we will comment on it below.
Following some familiar steps\cite{auerbach, fradkin}, 
we write the interacting fermion partition
function as an imaginary time coherent state path integral
\begin{equation}
Z = \int {\cal D}\big[\bar {\psi}(\tau)\;{\psi}(\tau)\big]\exp(-{\cal S})\;,
\label{partition_fnc}
\end{equation}
where the action ${\cal S}$ is
\begin{equation}
{\cal S} = \int_0^{\beta} d\tau\Big[\sum_{n,s}\bar{\psi}_{n,s}(\tau)
\Big({\partial \over \partial \tau} - \mu\Big){\psi}_{n,s}(\tau)
+ {\cal H}\big(\bar {\psi}(\tau), {\psi}(\tau)\big)\Big]\;.
\label{action}
\end{equation}
Here $\beta =1/k_{\rm B}T$, $\mu$ is the chemical potential
and we use units such that $\hbar=1$.
Since the exchange interaction term in our toy model is quadratic in the total
electron spin -- see Eq.~(\ref{toyhamiltonian}) -- 
its contribution to the action
for each time step $\tau_k = k\epsilon= k\beta/N$
can be represented by a Gaussian integral over a real vector field
$\vec \Delta_k \equiv \vec \Delta(\tau_k)$
\begin{widetext}
\begin{equation}
\int d\vec \Delta_k e^{-{{\cal N}_a\over 2U}\vec \Delta_k\cdot \vec \Delta_k +
\vec \Delta_k\cdot 
\frac{1}{2}
\sum_{n,s,s'}\bar\psi_{n,s}(\tau_k)\vec\sigma_{s,s'}\psi_{n,s'}(\tau_{k-1})}
\nonumber\\
\propto\; e^{-H_{\rm exch}\big(\bar\psi(\tau_k),\psi(\tau_{k-1}\big)}\;.
\end{equation}
Using this transformation for each time step, we obtain the following
functional integral over a single auxiliary field $\vec \Delta(\tau)$
that fluctuates in imaginary time
\begin{equation}
Z = \int {\cal D}\big[\bar {\psi}(\tau)\;{\psi}(\tau)\big]\;
\int {\cal D}\big[\vec \Delta(\tau)\big]\exp(-{\cal S})\;,
\label{partition_fnc2}
\end{equation}
where the action ${\cal S}$ is
\begin{eqnarray}
{\cal S}= 
\int_0^{\beta}& d\tau&\Big[ {{\cal N}_a \over 2U}
\vec \Delta(\tau)\cdot \vec \Delta(\tau)
+\sum_{n,s}\bar {\psi}_{n,s}(\tau)
\big({\partial \over \partial \tau} - \mu\big){\psi}_{n,s}(\tau)
+ H_{\rm band}\big(\bar {\psi}(\tau), {\psi}(\tau)\big)\nonumber\\  
&&+ H_{\rm so}\big(\bar {\psi}(\tau), {\psi}(\tau)\big) -
\frac{1}{2}
\big(\vec \Delta(\tau) + g_s\mu_{\rm B}\vec H_{\rm ext} \big)
\cdot\sum_{n,s}\bar{\psi}_{n,s}(\tau)\vec\sigma_{s,s'}{\psi}_{n,s}(\tau)\Big]\;.
\end{eqnarray}
This action can be written in another form which is especially useful
for small nanoparticles at low temperatures. 
Let us consider the fluctuating one-body Hamiltonian
\begin{equation}
\hat H_{1b}\big(\vec\Delta(\tau)\big) =
\hat H_{\rm band} + \hat H_{\rm so} -
\frac{1}{2}\big(\vec \Delta(\tau) + g_s\mu_{\rm B}\vec H_{\rm ext} \big)
\cdot\sum_{n,s}
c^{\dagger}_{n,s'}{\vec \sigma}_{s',s} c^{\phantom{\dagger}}_{n,s}\:.
\end{equation}
In Eq.~(\ref{partition_fnc2}) we can replace the functional integral 
over the fermionic coherent states
$\bar\psi(\tau), \psi(\tau)$ by a trace of the operator
\begin{equation}
\exp\Big(- \int_0^\beta d\tau \hat H_{1b}\big(\vec\Delta(\tau)\Big)=
\exp\Big(- \epsilon \sum_k\hat H_{1b}\big(\vec\Delta_k\big)\Big)
= \prod_k \exp\Big(- \epsilon \hat H_{1b}\big(\vec\Delta_k\big)\Big)\;.
\end{equation}
By inserting at each time step $\tau_k$ the resolution of the identity
$\hat I = \sum_a|\Psi_a(\vec\Delta_k)\big\rangle\langle\Psi_a(\vec\Delta_k)|$,
where $\{|\Psi_a(\vec\Delta_k)\rangle\}_a$ is a complete set of eigenstates
of the one-body Hamiltonian $\hat H_{1b}(\vec\Delta_k)$, and 
taking the $T \to 0$ limit, we obtain 
$Z= \int {\cal D}\big[\vec \Delta(\tau)\big]\exp(-{\cal S})$, with
\begin{eqnarray}
{\cal S} =&& \epsilon \sum_k  {{\cal N}_a \over 2U}\vec\Delta_k\cdot \vec\Delta_k
-\ln\Big[\langle\Psi_0(\vec\Delta_N)|
\exp\Big(-\epsilon\hat H_{1b}(\vec\Delta_{N-1})\Big) 
|\Psi_0(\vec\Delta_{N-1})\rangle\nonumber\\
&& \times \dots \times
\langle\Psi_0(\vec\Delta_1)|
\exp\Big(-\epsilon\hat H_{1b}(\vec\Delta_{N})\Big)
|\Psi_0(\vec\Delta_{N})\rangle\Big]\:.
\end{eqnarray}
Note that the fluctuating field $\vec\Delta(\tau)$ is bosonic
and satisfies periodic boundary conditions $\vec\Delta(0) =\vec\Delta(\beta)$.
Since $\epsilon \to 0$ we can rewrite the action in the form
\begin{eqnarray}
{\cal S} =&&\int_0^{\infty} d \tau\: \Bigg\{{{\cal N}_a \over 2U}
\vec\Delta(\tau)\cdot \vec\Delta(\tau)
-\frac{1}{\epsilon}\Big[\langle\Psi_0\big(\vec\Delta(\tau+\epsilon)\big)|
e^{-\epsilon\hat H_{1b}(\vec\Delta(\tau)}
|\Psi_0\big(\vec\Delta(\tau)\big)\rangle -1\Big]\Bigg\}\\
&&
=\int_0^{\infty} d \tau\;
\Bigg[ {{\cal N}_a \over 2U}\vec\Delta(\tau)\cdot \vec\Delta(\tau)
+ E_{1b}\big(\vec\Delta(\tau)\big)+
 \langle \Psi_0 \vert 
\partial \Psi_0/\partial \vec \Delta \rangle\cdot 
{\partial \vec\Delta(\tau) \over \partial\tau} \Bigg]\:.
\label{action_final}
\end{eqnarray}
\end{widetext}
Here  $E_{1b}\big(\vec\Delta(\tau)\big)$ is the ground-state
energy of $\hat H_{1b}\big(\vec\Delta(\tau)\big)$.   The approximations
that are implicit in these manipulations are valid when 
collective fluctuations of the spin-splitting field are not strongly
coupled to particle-hole excitations. 

Eq.~(\ref{action_final})
will play a crucial role in this paper.
The quantity:
\begin{equation}
E_{\rm tot}\big[\vec\Delta(\tau)\big]
 \equiv {{\cal N}_a \over 2U}\vec\Delta(\tau)\cdot \vec\Delta(\tau)
+ E_{1b}\big(\vec\Delta(\tau)\big)\;,
\end{equation}
gives the quantum energy functional of the magnetic nanoparticle 
as a function of
the fluctuating spin-splitting field $\vec\Delta(\tau)$; its classical
limit, obtained by evaluating at zero frequency (i.e. static
$\vec \Delta$ independent of time) $E_{\rm tot}\big[\vec\Delta(\tau)\big]$,  
corresponds to the phenomenological micromagnetic energy 
functional for a coherent magnetic particle with magnetostatic
contributions neglected.

The last term in the action
\begin{equation}
{\cal S}_{\rm Berry} \equiv \int_0^{\infty} d \tau\:
\langle \Psi_0 \vert
\partial \Psi_0/\partial \vec \Delta \rangle\cdot
{\partial \vec\Delta(\tau) \over \partial\tau}\;,
\label{berry1}
\end{equation}
is a Berry phase term, which is related to the reduction of the
total spin component along the magnetization axis due to
spin deviations from the ground state configuration\cite{niu1998, niu1999}.
As we will see in Sec.~(\ref{sec_gauss}),
the Berry phase contribution to the action captures
the quantization condition of the collective
elementary excitations and the way in which spin-orbit interactions affect
this quantization condition.

The Hubbard-Stratonovich decoupling can also be carried out 
in the case of the LDE model of Eq.~\ref{exchH}. Here 
we should in principle introduce an auxiliary field $\vec \Delta_{d,i}(\tau)$
that is dependent on the atomic site $i$.
Complicated inhomogeneous and noncollinear spin-splitting fields 
can indeed occur for very small
nanoparticles with a few tens of atoms\cite{car1998, aleks_SC2001}. 
In this paper however, we will
consider only a {\it coherent} {\it i.e.}  
site-independent spin-splitting field,
which is a good approximation for the relatively large nanoparticles 
(${\cal N}_a > 50$) that we are interested in.

\subsection{Mean-Field Approximation}
The mean-field approximation for the action is obtained by finding the 
value of the spin-splitting field at which it is minimized.  
Since the minimum occurs for a time-independent spin-splitting field
$\vec\Delta_{\rm MF}$, the 
Berry phase contribution to the action does not enter at this level.
In the coherent-field approximation, the mean-field spin-splitting satisfies:
\begin{eqnarray}
&& \frac{{\cal N}_a}{U} \vec \Delta_{\rm MF} + \frac{\partial E_{1b}(\vec \Delta)}{\partial\vec \Delta}\Big|_{\vec \Delta_{\rm MF}} \nonumber \\
&& = \frac{{\cal N}_a}{U} \vec \Delta_{\rm MF} - \langle 
\Psi_{MF} \vert \vec S_{\rm tot} \vert \Psi_{MF} \rangle =0\,,
\label{mfeq}
\end{eqnarray}
where $|\Psi_{\rm MF}\rangle \equiv \vert \Psi_0(\vec\Delta_{\rm MF})\rangle$.
$\vec \Delta_{\rm MF}$ may be determined either by minimizing 
the energy functional or by
solving the self-consistent-field equations implied by the second 
form of Eq.(~\ref{mfeq}).   The same set of mean-field equations can
be derived directly from the more general expression for the action 
in which the functional integral over fermion Grassmann variables is still 
present\cite{fradkin}.

\section{Gaussian Fluctuations}
\label{sec_gauss}
Our theory of elementary magnetic excitations of metallic nanoparticles
is based on a Gaussian fluctuation theory in which the action is expanded
to second order around the static mean-field values of the auxiliary fields. 
It will prove informative to contrast two approaches 
that can be used to evaluate the Gaussian fluctuation action.  
In the first approximation we work directly with the Berry phase 
and energy terms in the action, which are rather transparently 
related to micromagnetic theory and Landau-Liftshitz semi-classical
dynamics.  This approach cannot, however, deal directly with the 
coupling of collective and particle-hole elementary excitations.  
In the second approach we expand the action to second order around
$\Delta_{\rm MF}$, without introducing the quasi-static energy functional
as an intermediate step.  We will see that the two approaches
are equivalent in the limit of energies smaller than the minimum 
particle-hole excitation energy.

\subsection{Energy Functional Approach}
If we are interested only in low-energy excitations, which 
have slow dependence on imaginary time, we 
can approximate the energy function by its dependence on 
{\it static} field variations:
\begin{equation}
E_{\rm tot}(\vec \Delta) = E_{\rm MF} + \frac{1}{2} \tilde \Delta^{\alpha} 
{\Bigg [} \frac{{\cal N}_a \delta_{\alpha,\beta}}{U} + 
\frac{\partial^{2} E_{1b}(\vec \Delta)}
{\partial \Delta^{\alpha} \partial \Delta^{\beta}}\Bigg|_{\vec \Delta_{\rm MF}} 
{\Bigg ]} \tilde \Delta^{\beta}\;, 
\label{estaticexpansion}
\end{equation}
where 
$\tilde \Delta^{\alpha}\equiv \Delta^{\alpha} - \Delta_{\rm MF}^{\alpha}$,
and $E_{\rm MF}\equiv E_{\rm tot}(\vec \Delta_{\rm MF})$.
Since amplitude variations are energetically more costly, 
the dominant fluctuations
of the order parameter will be rotations.  
The second derivative represents the expansion of the micromagnetic
energy around its extremum; the dependence of energy on orientation 
is generally referred to as the anisotropy energy of the nanoparticle.

The bosonic field $\vec \Delta(\tau)$ is periodic in $[0,\beta]$.
Therefore the Berry phase contribution
to the action given in Eq.~(\ref{berry1}) can be rewritten as
the closed line integral that is path-dependent but not dynamics dependent: 
\begin{equation}
\label{berry_loop}
{\cal S}_{\rm Berry} = \oint d \vec \Delta \cdot \langle \Psi_0 \vert \partial \Psi_0/\partial \vec \Delta \rangle\:.
\label{berryphase}
\end{equation} 
Since
\begin{eqnarray}
\langle\Psi_0|{\partial\over \partial\vec \Delta} \Psi_0\rangle &+&
\langle\Psi_0|{\partial\over \partial\vec \Delta} \Psi_0\rangle^{\star}=
\langle\Psi_0|{\partial\over \partial\vec \Delta} \Psi_0\rangle +
\langle{\partial\over \partial\vec \Delta} \Psi_0|\Psi_0\rangle \nonumber\\
&=&{\partial\over \partial\vec \Delta}\langle\Psi_0|\Psi_0\rangle =0\;,
\end{eqnarray}
the Berry phase ${\cal S}_{\rm Berry}$ is pure imaginary.
For small amplitude rotations around the direction of $\Delta_{\rm MF}$, 
the closed line integral Eq.~(\ref{berry_loop})
is equal to the product of the area of the enclosed path 
and the Berry curvature,
\begin{equation}
{\cal C}(\Delta_{\rm MF}) =  \hat z \cdot \vec \nabla \times \langle \Psi_0 \vert \frac{\partial \Psi_0}{\partial \vec \Delta} \rangle
= \langle \frac{\partial \Psi_0}{\partial \Delta^{x}} \vert \frac{\partial \Psi_0}{\partial \Delta^{y}} \rangle - 
 \langle \frac{\partial \Psi_0}{\partial \Delta^{y}} \vert \frac{\partial \Psi_0}{\partial \Delta^{x}}\rangle.
\label{berrycurvature}
\end{equation}
Unlike the integrand of Eq.(~\ref{berryphase}), the Berry curvature 
is gauge independent, {\it i.e.} it is independent of 
the arbitrary phase choice made for the many-electron wavefunction at each
magnetization orientation.   
For small fluctuations the integrand in Eq.~(\ref{berryphase}) can be
chosen to be anything whose curl is the constant 
$C({\Delta_{\rm MF}}) \hat z$.  For example one convenient choice,
analogous to the symmetric gauge choice for a constant magnetic field, leads to  
\begin{equation}
{\cal S}_{\rm Berry} = \frac{{\cal C}(\Delta_{\rm MF})}{2} \oint 
\big[\hat z \times \vec \Delta\big]\cdot d \vec \Delta \; ,
\label{sberrygaussian} 
\end{equation}
if we choose the $\hat z$ direction to be the direction of $\vec \Delta_{\rm MF}$.
This line integral can be parametrized in terms of the imaginary time
variable
\begin{equation}
{\cal S}_{\rm Berry} = \frac{{\cal C}(\Delta_{\rm MF})}{2} \int_0^{\beta} d \tau 
\big[\tilde \Delta^{x} \partial_{\tau}\tilde \Delta^{y} -\tilde \Delta^{y} \partial_{\tau}\tilde \Delta^{x} \big]\;,
\label{tsberrygaussian} 
\end{equation}
which can also be written in frequency space as
\begin{eqnarray}
{\cal S}_{\rm Berry} =  \frac{1}{2\beta}\sum_{i\omega_n} 
&&i\omega_n\; {\cal C}(\Delta_{\rm MF})
\big [\tilde\Delta^x(-i\omega_n)\;\tilde\Delta^y(i\omega_n)\nonumber\\
&&-\tilde\Delta^y(-i\omega_n)\;\tilde\Delta^x(i\omega_n)\big]\;.
\label{wsberrygaussian}
\end{eqnarray}
Eq.~(\ref{wsberrygaussian}) shows that the Berry phase
contributes a term linear in frequency $\omega_n$ to the imaginary
part of the quadratic Gaussian fluctuation action. 
We will see that in the next section that the same contribution to the 
action arises from a small frequency expansion 
of the spin-fluctuation propagator kernel, defined in the next section.

\subsubsection{Calculation of the Berry's curvature}
\label{berry_tri}
We now briefly discuss the numerical evaluation of Berry curvatures \cite{niu1998} 
evaluated at the mean-field-solution point.
We start from the expression for ${\cal S}_{\rm Berry}$ given by Eq.~(\ref{berryphase}),
and consider paths on the unit sphere of exchange-field orientations. 
A small area closed path centered on the mean-field orientation 
encloses a small area on the unit sphere whose normal is the
direction of $\Delta_{\rm MF}$, which we take as the $\hat z$ direction.
Using Stoke's theorem and taking the Berry curvature out of the surface 
integral, we obtain  

\begin{equation}
{\cal S}_{\rm Berry} ={\cal C}(\Delta_{\rm MF})\int_{\rm Area}dS_{\Delta}\,,
\end{equation}\

For our calculation we consider as a closed path a small right  triangle
in the $xy$-plane of sides $\Delta_x=\Delta_y = \Delta_{\rm MF}\theta$, where
$\theta$ is a small angle describing the rotation around the $x$ and $y$ axis. 
Then

\begin{equation}
\label{triangle}
{\cal S}_{\rm Berry} =\frac{1}{2} {\cal C}(\Delta_{\rm MF})\Delta_{\rm MF}^2\theta^2\;.
\end{equation}\
This Berry phase can also be expressed as\cite{niu1998}

\begin{widetext}
\begin{equation}
i{\cal S}_{\rm Berry} = -{\rm Im}\,\ln\Big [\langle \Psi_{\rm MF}|\Psi_0(\vec\Delta_x)\rangle
\langle \Psi_0(\vec\Delta_x)|\Psi_0(\vec\Delta_y)
\rangle\langle \Psi_0(\vec\Delta_y)| \Psi_{\rm MF}\rangle\Big]\;,
\label{triangle2}
\end{equation}
\end{widetext}
where $|\Psi_0(\vec\Delta_x)\rangle$ and $|\Psi_0(\vec\Delta_y)\rangle$ are the
single Slater determinants of lowest energy for exchange fields
that differ from the mean-field value by 
$\Delta_x\hat x$, and $\Delta_y\hat y$ respectively.  Eq.~(\ref{triangle})
can be obtained by expanding $|\Psi_0(\vec\Delta_x)\rangle$ 
and $|\Psi_0(\vec\Delta_y)\rangle$ to quadratic order 
in $\Delta_x$ and $\Delta_y$,
remembering that 
$\langle\Psi_0|{\partial\over \partial\vec \Delta} \Psi_0\rangle$
is pure imaginary.
The expression for the Berry phase given in Eq.~(\ref{triangle}) is
very suitable for numerical calculations, since the wavefunctions
$|\Psi_0(\vec\Delta_x)\rangle$
and $|\Psi_0(\vec\Delta_y)\rangle$ can be easily calculated.
The arbitrary phases of the wave functions at the three vertices explicitly cancel
since each wave function {\em and} its complex conjugate appears in Eq.~(\ref{triangle}).
If we choose the arbitrary phases in such a way that the matrix elements
of $|\Psi_0(\vec\Delta_x)\rangle$ and $|\Psi_0(\vec\Delta_y)\rangle$
with $|\Psi_{\rm MF}\rangle$ are real and positive, 
then the Berry phase is given
by the simplified expression
\begin{equation}
i{\cal S}_{\rm Berry}= {\rm Im} 
\ln \langle \Psi_0(\vec\Delta_x)|\Psi_0(\vec\Delta_y) \rangle\;.
\end{equation}

Finally, one can show that the Berry phase (or Berry curvature) of the many-body
wave function is given by the sum of the Berry phases (or curvatures)
of the occupied single particle states, even in the presence of spin-orbit
interaction. The easiest way to prove this is to show that
\begin{equation}
\label{spberry}
\langle\Psi_0|{\partial\over \partial\vec \Delta} \Psi_0\rangle =
\sum_n^{\rm occ}\langle\phi_n|{\partial\over \partial\vec \Delta} 
\phi_n\rangle\,,
\end{equation}
where $\phi_n$ are single-particle eigenstates of the one-body Hamiltonian.
The gradient in Eq.~(\ref{spberry}) acts separately on the single-particle wave
functions and hence it can be regarded as a sum of single-particle operators.
It follows that the Berry phase can expressed as
\begin{equation}
i{\cal S}_{\rm Berry}= \sum_n^{\rm occ}{\rm Im} 
\ln \langle \phi_n(\vec\Delta_x)|  \phi_n(\vec\Delta_y)\rangle\;.
\end{equation}

\subsubsection{Berry phase without spin-orbit coupling}
For the case of no spin-orbit coupling, the Berry phase term can
be calculated analytically and is equal to the number of
singly-occupied orbitals times the usual spin-$1/2$ Berry phase.
To show this, let us consider a many-body wave-function within
the long-range exchange model describing a state polarized
in the  direction $\hat \Omega \equiv {\vec \Delta / \Delta}$ 
\begin{equation}
|\Psi_0(\vec \Delta) = \prod_s\big(u(\hat \Omega)c^{\dagger}_{s\uparrow} +
v(\hat \Omega)c^{\dagger}_{s\downarrow}\big)\,
\prod _d c^{\dagger}_{d\uparrow} c^{\dagger}_{d\downarrow}|0\rangle
\equiv |\Psi_0(\hat \Omega)\;,
\end{equation}
where the index $s$ runs over the $N_{\uparrow} - N_{\downarrow}$ 
singly-occupied states and
the index $d$ over the $N_{\downarrow}$ doubly-occupied states.
The functions $u(\hat \Omega)$ and $v(\hat \Omega)$ 
are written in terms of the angles $\Theta$ and $\Phi$ specifying
the unit vector $\hat \Omega(\Theta,\Phi)$
\begin{eqnarray}
u(\hat \Omega)=\cos(\Theta/2)\,,\\
v(\hat \Omega)=e^{i\Phi}\sin(\Theta/2)\,.
\end{eqnarray}
We obtain 
\begin{eqnarray}
&&\big |{\partial \Psi_0(\hat \Omega)\over \partial \hat \Omega}\big \rangle=
\sum_s\big({\partial u \over  \partial \hat \Omega}c^{\dagger}_{s\uparrow}
+ {\partial v \over  \partial \hat \Omega}c^{\dagger}_{s\downarrow}\big)
\nonumber\\
&&\times\prod_{s'\neq s}\big(u(\hat \Omega)c^{\dagger}_{s\uparrow} +
v(\hat \Omega)c^{\dagger}_{s\downarrow}\big)\,
\prod _d c^{\dagger}_{d\uparrow} c^{\dagger}_{d\downarrow}|0\rangle\,.
\end{eqnarray}
Hence, for rotations of the spin-splitting field
\begin{eqnarray}
&&\langle\Psi_0|{\partial\Psi_0\over \partial\vec \Delta}\rangle = 
\langle\Psi_0|{\partial \Psi_0\over \partial\hat \Omega}\rangle =
\sum_s\big( u^{\star} {\partial u \over  \partial \hat \Omega}
+ v^{\star}{\partial v \over  \partial \hat \Omega}\big) \nonumber\\
&=& \sum_s  i\;{1-\cos(\Theta)\over 2\sin(\Theta)}\hat \Phi
= i\,{N_{\uparrow} - N_{\downarrow}\over 2}\,
{1-\cos(\Theta)\over \sin(\Theta)}\,\hat \Phi\;,
\end{eqnarray}
and the Berry phase
\begin{eqnarray}
{\cal S}_{\rm Berry}&&= \int_0^{\beta}d\, \tau 
\langle\Psi_0|{\partial \Psi_0\over \partial\hat \Omega}\rangle
\cdot {d\,\hat \Omega\over d\, \tau}
=\oint \langle\Psi_0|{\partial \Psi_0\over \partial\hat \Omega}\rangle
 \cdot d\,\hat \Omega\nonumber\\
&=& i\, {N_{\uparrow} - N_{\downarrow}\over 2}\, 
\oint d \Phi \big(1- \cos(\Theta_{\Phi})\big)\nonumber\\
&=& i\, \big[{N_{\uparrow} - N_{\downarrow}}\big] {A[\hat \Omega]\over 2}\;,
\end{eqnarray}
where $A[\hat \Omega]$ is the area enclosed 
by the path $\Omega(\tau)$ on the
unit sphere, and
$A[\hat \Omega]/2$ is the usual spin-$1/2$ Berry phase.  In the presence of 
spin-orbit interactions, the Berry phase becomes highly non-trivial 
as we discuss in Sec.\ref{numerics}.

For a path enclosing the small right triangle described
in Sec.~\ref{berry_tri}, $A[\hat \Omega] = \theta^2/2$ and
\begin{eqnarray}
\label{berry_red}
{\cal S}_{\rm Berry} &=& i {\theta^2 S\over 2}\;,\\
{\cal C}(\Delta_{\rm MF})&=&i {S\over \Delta_{\rm MF}^2}\;,
\label{berryC}
\end{eqnarray}
where $S= (N_{\uparrow} - N_{\downarrow})/2$ is the
total spin in the nanoparticle ground state.
\subsection{Gaussian Fluctuations: Perturbation Theory Approach}
The Gaussian fluctuations theory can be derived more generally by 
formally integrating out the fermions and then expanding second order around $\Delta_{\rm MF}$
without making any quasi-static approximations.  We discuss the case of the long-range
exchange interaction first and then indicate what changes occur in the short-range (d-only) exchange model.
The formally exact expression for the action is
\begin{equation}
S = \int_{0}^{\beta} \frac{{\cal N}_a}{2U} {\vec \Delta} \cdot {\vec \Delta} - 
\ln \det (\partial_{\tau} + H_{1b}(\Delta_{\rm MF}^{\alpha} + \tilde \Delta^{\alpha})).
\label{formalaction}
\end{equation}
where $H_{1b}(\Delta)$ includes, in addition to its single-particle hopping and spin-orbit terms, the spin splitting terms $ -(\vec \Delta_{\rm MF} + 
\vec H_{\rm ext}) \cdot \vec s
+ \tilde \Delta^{\alpha} s^{\alpha}$.  It is the second of these two terms that is treated 
perturbatively.  First order terms in the expansion vanish, since the mean-field value of 
$\vec \Delta$ is an extremum of the action.  
The second order terms can be obtained by a standard
calculation with the following result:
\begin{equation} 
{\cal S}_{\rm fluc} = \frac{1}{2 \beta} \sum_{i\omega_n} \tilde \Delta^{\beta}(-i \omega_n) 
K_{\beta\,\alpha}(i\omega_n) \tilde \Delta^{\alpha}(i\omega_n)\;,
\label{sflucpt}
\end{equation}
where the kernel $K_{\beta\,\alpha}(i\omega_n)$ is the inverse 
of the exchange-field-fluctuation propagator and is given by 
\begin{eqnarray}
K_{\beta\,\alpha}(i\omega_n) = \delta_{\alpha,\beta} 
\frac{{\cal N}_a}{U} 
+ &\sum_{I,J}& \frac{n_F(\xi_J) - n_{F}(\xi_I)}{i \omega_n 
+ \xi_J - \xi_I}\nonumber\\ 
&&\times\langle J \vert s^{\beta} \vert I \rangle 
\langle I \vert s^{\alpha} \vert J \rangle\;,
\label{flucprop}
\end{eqnarray}
where $\vert I \rangle$ is a mean-field electron eigenstate and $\xi_{I}$ is the corresponding eigenvalue with 
energies measured from the chemical potential.  

Fluctuations in the magnetization orientation around the mean field
direction $\hat z =\vec \Delta_{\rm MF} / \Delta_{\rm MF}$, are described
by the transverse diagonal $K_{x\,x}$, $K_{y\,y}$ and off-diagonal
$K_{x\,y}$, $K_{y\,x}$ components of the kernel. It is customary to introduce
$K_{+\,-}(i\omega_n)$ and $K_{-\,+}(i\omega_n)$.
When the spin-orbit interaction is present, $K_{+\,+}(i\omega_n)$ 
and $K_{-\,-}(i\omega_n)$ 
do not vanish and play a role. Note that 
\begin{equation}
\label{kpm}
K_{-\,+}(i\omega_n)=K_{+\,-}(-i\omega_n)\,,
\end{equation}
\begin{equation}
\label{kpp}
K_{+\,+}(-i\omega_n)=K_{+\,+}(i\omega_n)\,,
\end{equation}
\begin{equation}
\label{kmm}
K_{-\,-}(i\omega_n)=K_{+\,+}(-i\omega_n)^{\star}\,.
\end{equation}
It follows form Eq.~(\ref{flucprop}) that
\begin{equation}
\label{kxx}
K_{x\,x} = {\cal N}_a/U + (K_{+\,-}+K_{-\,+})/4 + (K_{+\,+}+K_{-\,-})/4\,, 
\end{equation}
\begin{equation}
\label{kyy}
K_{y\,y} =  {\cal N}_a/U + (K_{+\,-}+K_{-\,+})/4 - (K_{+\,+}+K_{-\,-})/4\,, 
\end{equation}
\begin{equation}
\label{kxy}
K_{x\,y}=  i (K_{+\,-}-K_{-\,+})/4 - i(K_{+\,+}-K_{-\,-})/4\,,
\end{equation}
and
\begin{equation}
\label{kyx}
K_{y\,x}= -i (K_{+\,-}-K_{-\,+})/4 - i(K_{+\,+}-K_{-\,-})/4\,.
\end{equation}

The transverse fluctuation action of the long-range exchange model 
is quite simple in the absence of spin-orbit interactions.
Firstly, the components $K_{+\,+}=K_{-\,-}=0$, since the spin of
the quasi-particle states is a good quantum number.  (This property 
depends only on spin-rotational invariance and holds for the short-range exchange model as well). 
Furthermore, since the 
the electron mean-field eigenstates factorize into spin and orbital factors  
identical spin-up and spin-down wavefunctions are 
split energetically by $\Delta_{\rm MF}$ and we obtain
\begin{equation}
K_{+\,-}(i \omega_n) = 
\frac{N_{\uparrow} - N_{\downarrow}}{i \omega_n -(\Delta_{\rm MF} + 
H_{\rm ext})}
= K_{-\,+}(-i \omega_n)\,.
\label{nospinorbit}
\end{equation}
In examining the consequences of this simple property for the Gaussian action kernel,
we first consider $K_{x\,y}(i\omega_n)$. 
We see from Eq.~(\ref{nospinorbit})
that when there is no spin-orbit interaction, $K_{x\,y}(i\omega_n =0) =0$,
a property that reflects invariance under rotations of exchange-field orientation around the $\hat z$ axis.
By expanding $K_{x\,y}(i\omega_n)$ around $i\omega_n=0$, we obtain
\begin{equation}
K_{x\,y}(i\omega_n)\approx -i\omega_n\, i\,{N_{\uparrow} - N_{\downarrow}
\over 2\Delta_{\rm MF}^2}
=  -{i\, S\over  \Delta_{\rm MF}^2}\,i\omega_n\;.
\label{kxy_w}
\end{equation}
By inserting 
Eq.~(\ref{kxy_w}) into Eq.~(\ref{sflucpt}) and comparing the result
with Eq.~(\ref{wsberrygaussian}), 
we can identify the term of $K_{x\,y}(i\omega_n)$ linear in $i\omega_n$,
$i\,{S/ \Delta_{\rm MF}^2}$, with the Berry curvature,
${\cal C}(\Delta_{\rm MF})$, in agreement with Eq.~(\ref{berryC}).

Similarly, the expansion of $K_{x\,x}(i\omega_n) =K_{y\,y}(i\omega_n)$
around $i\omega_n=0$ yields
\begin{eqnarray}
K_{x\,x}(i\omega_n)=K_{y\,y}(i\omega_n)&\approx&
{{\cal N}_a\over U} - {N_{\uparrow} - N_{\downarrow}\over 2\Delta_{\rm MF}}
+ {\cal O}(\omega_n ^2)\nonumber\\ 
&=&0 + {\cal O}(\omega_n ^2)\,,
\label{kxx_w}
\end{eqnarray}
where in the last equality of Eq.~(\ref{kxx_w})
we have used the mean-field relationship $(U/{\cal N}_a)\, S = \Delta_{\rm MF}$
between the total magnetization of the nanoparticle and the spin-splitting
field. The vanishing of the diagonal components $K_{x\,x}$ and $K_{y\,y}$
at $\omega_n =0$ is again a result of rotational invariance.  In the absence of 
spin-orbit interactions there is a collective mode at zero energy because of 
magnetization-orientation rotational invariance.

The physics that we wish to investigate in this paper is largely contained in the way that the simple
results outlined above are altered by spin-orbit interactions. 
As we have stated, both Berry-phase and the energy-term in the action become highly non-trivial,  
a property that we now examine from the perturbation theory point-of-view.
From Eqs.~(\ref{kpm})-(\ref{kyx})
it is clear that $K_{x\, x}(\omega_n)$, $K_{x\, x}(\omega_n)$, 
$K_{x\, y}(\omega_n)$ and $K_{x\, y}(\omega)$
are real at $\omega_n=0$.  Comparing Eq.~(\ref{estaticexpansion}) 
with the $\omega_n=0$ term of
Eq.~(\ref{sflucpt}), we see that the kernel coefficients reduce to the time-independent perturbation theory 
expressions for the second derivatives of the total energy with respect to exchange field, 
$\partial^2 E_{\rm tot}(\vec \Delta_{\rm MF})/
\partial \Delta^{x}\,\partial \Delta^{x}$, 
$ \partial^2 E_{\rm tot}(\vec \Delta_{\rm MF})/
\Delta^{y}\,\partial \Delta^{y}$.
In its static limit, the fluctuation kernel reduces to one that would be obtained from
a classical theory with the micromagnetic energy functional derived from a mean-field-theory
calculation, or for accurate first principles calculations by solving spin-density-functional
Kohn Sham equations self-consistently.
By expanding $K_{\alpha\, \beta}(i\omega_n)$ around $i\omega_n=0$,
we obtain
\begin{widetext}
\begin{eqnarray}
\label{kxxw0}
K_{x\, x}(i\omega_n) &\approx & {\partial^2 E_{\rm tot}(\vec \Delta_{\rm MF})
\over \partial \Delta^{x}\,\partial \Delta^{x}}+{\cal O}(\omega_n^2)=
(a+ b) +{\cal O}(\omega_n^2)\:, \\
\label{kyyw0}
K_{y\, y}(i\omega_n) &\approx & {\partial^2 E_{\rm tot}(\vec \Delta_{\rm MF})
\over \partial \Delta^{y}\,\partial \Delta^{y}}+{\cal O}(\omega_n^2)
=(a- b) +{\cal O}(\omega_n^2)\:, \\
\label{kxyw0}
K_{x\, y}(i\omega_n) &\approx& {\partial^2 E_{\rm tot}(\vec \Delta_{\rm MF})
\over \partial \Delta^{x}\,\partial \Delta^{y}} - 
{\cal C}(\Delta_{\rm MF})\; i\omega_n
=c - {\cal C}(\Delta_{\rm MF})\; i\omega_n
\;,\\ 
\label{kyxw0}
K_{y\, x}(i\omega_n) &\approx& {\partial^2 E_{\rm tot}(\vec \Delta_{\rm MF})
\over \partial \Delta^{y}\,\partial \Delta^{y}}+
{\cal C}(\Delta_{\rm MF})\; i\omega_n 
= c + {\cal C}(\Delta_{\rm MF})\; i\omega_n
\;,
\end{eqnarray}
\end{widetext}
where
\begin{equation}
\label{a_real}
a ={{\cal N}_a\over U}+{K_{+\, -}(0)\over 2}\;, 
\end{equation}
\begin{equation}
b={K_{+\, +}(0)+K_{+\, +}(0)^{\star}\over 4}\;,
\end{equation}
\begin{equation}
c=-i{K_{+\, +}(0)-K_{+\, +}(0)^{\star}\over 4}\;,
\end{equation}
are real constants and the Berry curvature is
\begin{equation}
\label{berry_curvPT}
{\cal C}(\Delta_{\rm MF}) = {i\over 2} 
\sum_{I,J} \frac{n_F(\xi_J) - n_{F}(\xi_I)}{(\xi_J - \xi_I)^2} 
\big |\langle J \vert s^{+} \vert I \rangle\big|^2\; .
\end{equation}
This expression for the Berry curvature can also be derived from 
Eq.(~\ref{berrycurvature}) by using time-independent perturbation
theory expressions for the dependence of single-particle wavefunctions
on exchange-field orientation. 
Note that the leading frequency dependences in $K_{x\, x}$ and $K_{y\, y}$ 
are quadratic.
These equations are valid for $i\omega_n$ smaller
than the smallest particle-hole excitation energy $|\xi_J - \xi_I| $,
as we discuss at greater length below.
So far we have completely disregarded amplitude fluctuations of the 
exchange field,{\it i.e.} the components of the Kernel involving 
$\Delta^{z}$.  It is obvious that in absence of
spin-orbit interactions
\begin{eqnarray}
K_{x\; z}(i\omega_n)&=&K_{z\; x}(i\omega_n)=0\;,\\
K_{y\; z}(i\omega_n)&=&K_{z\; y}(i\omega_n)=0\;,\\
K_{z\; z}(i\omega_n)&=& {{\cal N}_a\over U}
+ \frac{\delta_{n,0}}{4}\sum_I 
{\partial n_{\rm F}\over \partial \xi}\Bigg|_{\xi=\xi_I}\;.
\end{eqnarray}
When spin-orbit is present, the off-diagonal components involving
$\Delta^{z}$ are in principle non-zero. 
However it turns out that they are always very small 
(see next section); therefore  we will typically neglect them and we will only
keep $K_{z\; z}(i\omega_n)={{\cal N}_a/ U}$.  This approximation amounts to 
neglecting the weak dependence of the magnitude of the exchange field on
its orientation. 

%

\subsection{Excitation Spectrum}
\label{spectrum}

The kernel of the fluctuation action is the {\it matrix inverse}
of the exchange-field propagator, which is proportional to the spin susceptibility
linear response function.  We refer to this propagator below as the spin-susceptibility.
\begin{equation}
\chi_{\alpha\,\beta}(i\omega_n) = [K(i\omega_n)^{-1}]_{\alpha\, \beta}
\label{susceptibility}
\end{equation}
The elementary magnetic excitations of the nanoparticle (particle-hole and 
collective) occur at real frequencies frequencies $\omega$
at which $\chi_{\alpha\,\beta}(\omega)$ has poles. 
Here $\chi_{\alpha\,\beta}(\omega)$ is obtained by analytically
continuing Eq.~(\ref{susceptibility}) to real frequencies: 
$i\omega_n \to \omega + i\eta$. Let $D(\omega)$ denote the complex
function,
\begin{equation}
D(\omega)= {\rm det}\big[ K(\omega) \big]\;.
\end{equation}
The condition
for the existence of a pole in $\chi_{\alpha\,\beta}(\omega)$ can
be expressed by the equation
\begin{equation}
D(\omega)= 0
\end{equation}
In order to make progress, we will assume that all the components
of $K$ involving the longitudinal variable $z$ are small and
can be disregarded, except for $K_{z\, z}(\omega)$
The expression for the elementary excitation energy then simplifies to
\begin{eqnarray}
D(\omega)
\approx K_{z\, z}\Big[K_{x\, x}(\omega)\,K_{y\, y}(\omega) -
K_{x\, y}(\omega)\,K_{y\, x}(\omega)\Big]=0\; ,
\label{det0}
\end{eqnarray}
where we have used Eqs.~(\ref{kxx}, \ref{kxy}).

Let us consider first the limit of small nanoparticles,
where the single-particle mean-level spacing $\delta$ is much larger than the 
ferromagnetic resonance, which is of order the anisotropy energy per atom, $K$.
In Cobalt $K \approx 0.1 $ meV, and if use bulk 
single-particle density of states\cite{papaconstantopoulos}
we find that this limit is reached in nanoparticles containing
$N_a << 10^4$ atoms. In this case we expect a pure 
ferromagnetic resonance mode with a large spectral weight that 
appears as a separate quantum state below the lowest particle-hole
excitation energy. (We discuss the situation where particle-hole and 
collective excitations are not cleanly separated at greater length 
below.)  
Using the expansions of Eqs.~(\ref{kxxw0}--\ref{kxyw0}), we obtain
\begin{equation}
{{\cal N}_a\over U} 
\Big[a^2 - b^2 - c^2 - \big(i\,{\cal C}(\Delta_{\rm MF})\big)^2\,
\omega^2\Big]=0\,, 
\label{omega0}
\end{equation}
which yields a low-energy pole at the real frequency
\begin{equation}
E_{\rm res} = {\sqrt{a^2 - b^2 - c^2}
\over 
\big|i{\cal C}(\Delta_{\rm MF})\big|}\,. 
\label{wres}
\end{equation}
Around the pole we have,

\begin{equation}
 \big[D(\omega) \big]^{-1}
 \approx {{\cal Z}_{\rm res}\over E_{\rm res} - \omega}\:,
\end{equation}
where the ``residue'' ${\cal Z}_{\rm res}$ is

\begin{equation}
{\cal Z}_{\rm res}^{-1} = 2\,{N_a\over U}\,
\sqrt{a^2 - b^2 - c^2}\,
\big|i{\cal C}(\Delta_{\rm MF})\big|\;.
\label{resZ}
\end{equation}
Thus the collective excitations have a gap at 
$E_{\rm res}$. The gap is proportional 
to the {\it quotient}
of $\sqrt{a^2-b^2-c^2}$, which is essentially the total anisotropy energy,
and the Berry curvature ${\cal C}(\Delta_{\rm MF})$. For particles that
are not too small, both quantities are proportional to the particle volume,
thus $E_{\rm res}$ is approximately independent of particle volume
and approximately equal to the anisotropy energy/atom, $K$.
The spectral weight of the collective mode
${\cal Z_{\rm res}}$ is inversely proportional to the {\it product} of the same
two quantities.  The spectral weight divided by the resonance frequency 
is proportional to the static response of the exchange-field, a quantity 
which can be understood simply by minimizing the total micromagnetic energy.

The considerations described above apply when the collective excitation energy
is smaller than the lowest energy particle-hole excitation.  Even in 
this limit where particle-hole and collective excitations are well
separated, the exchange-field propagator does have some spectral weight in 
particle hole excitations as well.  The Kernel (Eq.~\ref{flucprop})
must have zeroes between the poles that occur at each mean-field
particle-hole excitation energy.  Therefore we expect $D(\omega)=0$ at 
$\omega=\tilde \omega_{ji}\approx \omega_{ji} = \xi_{I} - \xi_{J}$
and the susceptibility to have additional poles
at $\tilde \omega_{ji}$. Let us expand $D(\omega)$ around one of these poles
\begin{equation}
\big[D(\omega)\big]^{-1}\approx \, 
{\big[D'(\tilde \omega_{ji})]^{-1}\over  \omega -  \tilde \omega_{ji}}\;.
\label{det_exp}
\end{equation}
Thus, in general the residue of each pole is proportional to 
\begin{equation}
{\cal Z}_{ji} =  \big[D'(\tilde \omega_{ji})]^{-1}\,. 
\label{res_finite_w}
\end{equation}

\subsection{Coupling between Collective and Particle-hole excitations} 

If we increase the size of the nanoparticle to a few thousands atoms, the lowest
particle-hole excitation energy $\omega_{ji}\approx \delta$
will start to approach $E_{\rm res}$
from above. In this situation the pole at $E_{\rm res}$ will start
to loose its collective character.
%
As $\delta$ becomes much smaller
than $E_{\rm res}$, the ferromagnetic resonance mode 
will no longer be a single excitation, 
and instead appear as enhanced spectral weight in response functions 
that is spread over several particle-hole excitations. 
If we further increase the nanoparticle size such that 
$\delta << E_{\rm res}$, we start to approach the thermodynamic limit,
where the collective ferromagnetic resonance
will be spread over a large number of
particle-hole excitations. If $E_{\rm res}$ is the frequency
at which Eq.~\ref{det0} holds, by expanding $D(\omega)$ near $E_{\rm res}$
the ferromagnetic collective mode has a width given by
\begin{equation}  
\Gamma = 
2 {{\rm Im}\; D(E_{\rm res})\over {\rm Re}\; D'(E_{\rm res})}\;.
\end{equation}
The susceptibility then has a resonant denominator and ferromagnetic
mode has the shape of an asymmetric Lorentzian\cite{callaway1983}.

It is possible to relate the typical value of matrix elements between
single-particle states that appear in the kernel for a {\it large} nanoparticle
and the Gilbert damping parameter usually used to characterize the 
width of a ferromagnetic resonance line.  The Gilbert damping 
parameter $\alpha$ is the ratio between the line-width and the resonance
frequency.  It is normally introduced as a phenomenological damping parameter 
in Landau-Liftshitz equations of motion for the magnetization direction; 
these equations are implied by the low-frequency dynamics discussed above.
Taking the continuum limit for the particle-hole excitation 
spectrum and
assuming that we have approximate rotational symmetry about
the exchange-field-orientation axis, 
it follows from the preceding analysis that 
\footnote{In the continuum limit the imaginary part of $K_{xx}$ and $K_{yy}$ 
will have terms linear in $\omega$, which cause 
the ferromagnetic resonance mode to have a finite lifetime.}
\begin{equation}
\alpha = \frac{{\rm Im}[K_{xx}(\omega=E_{\rm res})]}
{{\rm Im}[K_{xy}(\omega=E_{\rm res})]}
\sim  \frac{\pi }{2 \delta^2 |{\cal C}(\Delta_{\rm MF})|}
\overline{ |\langle J | s_x | I \rangle |^2 } 
\label{alphaestimate}
\end{equation}   
where the overbar denotes an average over typical particle-hole matrix elements
for states near the Fermi energy.  Note that $\alpha$ approaches a
constant and the typical matrix element scales like ${\cal N}_A^{-1}$ in the 
limit of large nanoparticles.  

We can estimate the number of individual particle-hole excitations
that contribute to the ferromagnetic resonance in a large nanoparticle.
The density-of-states for particle-hole excitations grows linearly
with energy and at the resonance energy is $\sim E_{\rm res}/\delta^2$.
The number of particle-hole excitations in the resonance is 
the width of the resonance $\alpha E_{\rm res}$ times this density of states,
$\sim \alpha E_{\rm res}^2/\delta^2$.  This expression implies a condition
for the crossover to pure-state ferromagnetic resonance when 
$\delta = \sqrt{\alpha}\; E_{\rm res}$.  The crossover is therefore expected to occur 
at larger particle sizes when the ferromagnetic resonance is sharp.
Even when the particle is small enough that the resonance is not 
normally coupled with particle-hole excitations, avoided crossings between
these two types of excitations will frequently occur as the external
magnetic field or other parameters are varied.  We can describe such an
avoided crossing by assuming that the low-frequency limit can be taken
for all but one of the particle-hole excitation contributions to the kernel.
To briefly explain what happens in this limit, 
for which the kernel can be written as
the sum of a part $K^{\rm smooth}_{xx}$ 
with a smooth frequency dependence and a part  $K^{\rm res}_{xx}$ with a 
resonant frequency dependence, 
we assume that the particular 
particle-hole excitation contributes only to $K^{\rm res}_{xx}$, 
making a contribution
\begin{equation}
K^{\rm res}_{xx}(\omega) = \frac{2 \omega_{ij}}{\omega^2 - \omega_{ij}^2} 
|\langle J | s_x | I \rangle |^2.
\label{deltakxxqp}
\end{equation}
It follows that the poles of the exchange-field propagator occur at energies
\begin{equation}
\omega_{\pm} = \frac{E_{\rm res}+\omega_{ij}}{2} \pm
\big[ [(E_{\rm res}-\omega_{ij})/2]^2 + V^2 \big]^{1/2}
\label{energies}
\end{equation} 
where $V$ is an avoided crossing gap.  Using Eq.(~\ref{alphaestimate}) we find
that $V \sim \alpha E_{\rm res}$.  
The size of the avoided crossing gap that occurs in
a small nanoparticle when a particle-hole excitation energy is tuned 
through a collective excitation energy by an external field or another parameter is 
specified by the ferromagnetic resonance width in the limit of a large particle.

It is important to note that 
the experiments in Ref.~\onlinecite{gueron1999,deshmukh2001}
are carried out for nanoparticles containing ${\cal N}_a \approx 1500$ atoms.
Thus these experiments are probing the most interesting and {\it difficult}
intermediate regime of particle size,
where the lowest particle-hole excitation energy 
$\omega_{ji} \approx \delta \to \sqrt{\alpha} E_{\rm res}$.

\section{Numerical Results and Discussion}
\label{numerics}

%
In this session we will present numerical calculations performed
on our two ferromagnetic nanoparticle models which 
illustrate and support all the main points of the theory of elementary
spin excitations developed above. 
Since we can deal with nanoparticles containing 
up to 260 atoms, we expect, on the basis of our theoretical considerations,
that the exchange-field correlation function will normally display one main peak
at energies below the lowest particle-hole excitation. In addition to
verifying this expectation, we will see that the ferromagnetic resonance mode
is characterized by a very interesting behavior when manipulated with
an external magnetic field to bring it through con incidence
with a particle-hole excitation.

As mentioned already in Sec.~\ref{sec_gauss},  
in discussing the different components of the kernel
and the susceptibility we will take the $z$-direction along the direction
of the magnetization, $\hat \Omega$.
We will also choose the $y$-direction to lie in the equatorial $XY$-plane of 
the nanoparticle. Then the  $x$-direction will be fixed by the
condition of being orthogonal to both $z$ and $y$. Therefore
the $x$ and $y$ components of the kernel describe
{\it transverse} spin fluctuations with respect to the direction
of the magnetization, but these directions in general
will not be symmetry directions
with respect the crystal structure or the nanoparticle geometry. 
The direction of $\hat \Omega$ is determined by minimizing the
the classical micromagnetic energy functional
$E_{\rm tot}(\vec{\Delta})$, or equivalently by solving the 
mean-field equations self-consistently.  The variation of $E_{\rm tot}(\vec{\Delta})$
as a function of $\Theta$ and $\Phi$ 
has been studied in detail in 
Ref.~\onlinecite{ac_cmc_ahm2002}. 
 \begin{figure}
 \includegraphics{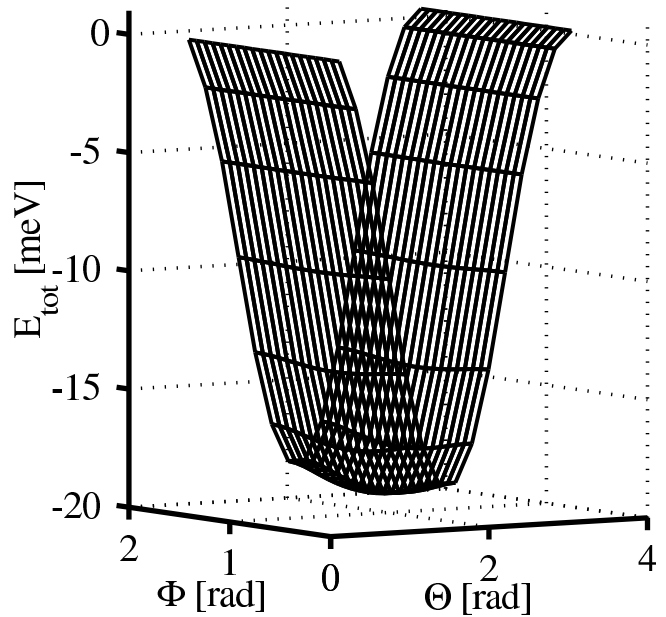}
 \caption{Anisotropy landscape $E_{\rm tot}(\vec{\Delta})$
as a  function of $\Theta$ and $\Phi$ for a 143-atom hemispherical nanoparticle.
$E_{\rm tot}$ is periodic in $\Phi$, with period $\pi/2$.}
 \label{anisotropy}
 \end{figure}
The main conclusions are summarized
in Fig.~\ref{anisotropy}, where we plot $E_{\rm tot}(\vec{\Delta})$ in the 
$(\Theta,\Phi)$-plane at $\vec H_{\rm ext} =0$ 
for a hemispherical nanoparticle consisting 
of ${\cal N}_A= 143$ Co atoms, arranged in a f.c.c. lattice.
In this case the magnetization
direction $\hat \Omega$ lies in the XY-plane ($\Theta= \pi/2$),
along one of the four directions
corresponding to the four degenerate shallow minima of 
$E_{\rm tot}(\vec{\Delta})$.
By applying an external field some of these minima will become classically
metastable and we can have different hysteretic behaviors with zero, one
or two reversals of the magnetic moment, depending on the direction of
the field.

\subsection{Real part of $K_{x\, x}(\omega)$}

 \begin{figure}
 \includegraphics{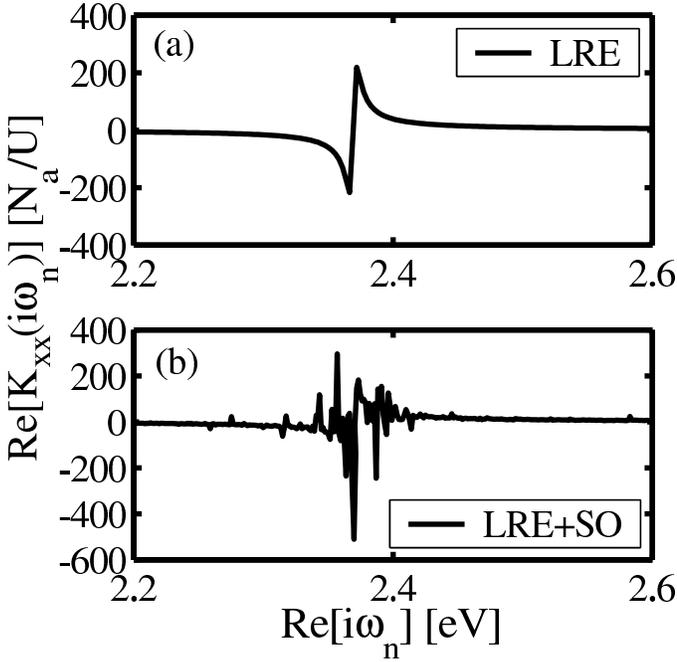}
 \caption{Real part of the transverse diagonal component of the kernel for
a hemispherical 26-atom nanoparticle, Long-Range Exchange model. 
(a) Without spin-orbit (SO) interaction. 
(b) With SO interaction.}
 \label{fig1_zeemanKernel}
 \end{figure}

\begin{figure}
 \includegraphics{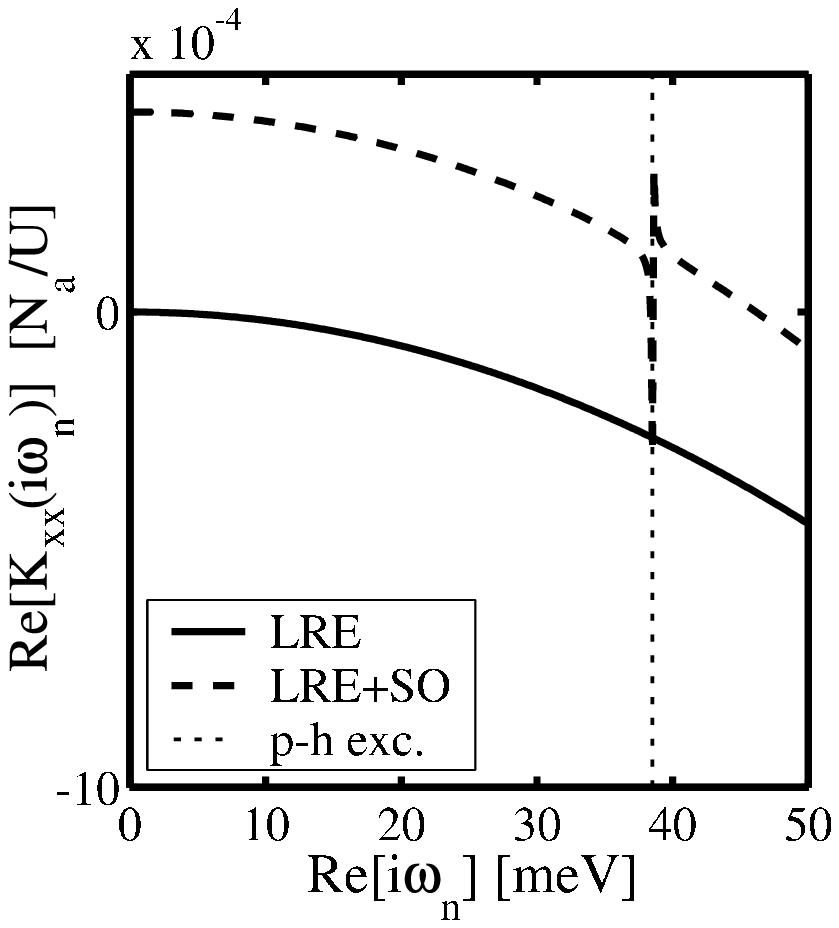}
 \caption{Low-energy behavior of the Kernel for the same system as 
in Fig.~\ref{fig1_zeemanKernel}. 
Solid line: without SO interaction; dashed line:
with SO interaction. The vertical dotted line marks the position of the
first particle-hole excitation, where ${\rm Re} K_{x\, x}$ has a pole.} 
 \label{fig2_zeemanKernelzoom}
 \end{figure}

We first consider the spectral representation of the kernel defined in
Eq.~(\ref{flucprop}). In Fig.~\ref{fig1_zeemanKernel}(a) we plot 
${\rm Re} K_{x\, x}(i\omega_n)$ vs. ${\rm Re}(i\omega_n)$
-- after analytical continuation $i\omega_n \to \omega + i\eta$ --  
for a 26-atom nanoparticle for the case of the LRE model. 
Consistent with
Eqs.~(\ref{nospinorbit}, \ref{kxx_w}), for zero SO 
we find that the Kernel has just one pole
at particle-hole excitation energy equal $\Delta_{\rm MF}$,
and a gapless zero. With SO interaction included, many more
matrix elements $\langle J \vert s^{+} \vert I\rangle$
will be non-zero, and consequently many more poles corresponding to
these particle-hole excitations will appear in the kernel. This is
shown in Fig.~\ref{fig1_zeemanKernel}(b), where we can see 
that the poles with largest residues still correspond to
particle-hole excitations near $\Delta_{\rm MF}$. Although
not visible in Fig.~\ref{fig1_zeemanKernel}(b) because of its very small 
residue, the first pole in the kernel occurs at an energy of
order of the single-particle mean level spacing $\delta$, 
which for this particle size
is approximately $20\ {\rm meV}$. 
This is shown in Fig.~\ref{fig2_zeemanKernelzoom}, where we zoom
on a small energy window at very low energies.

 \begin{figure}
 \includegraphics{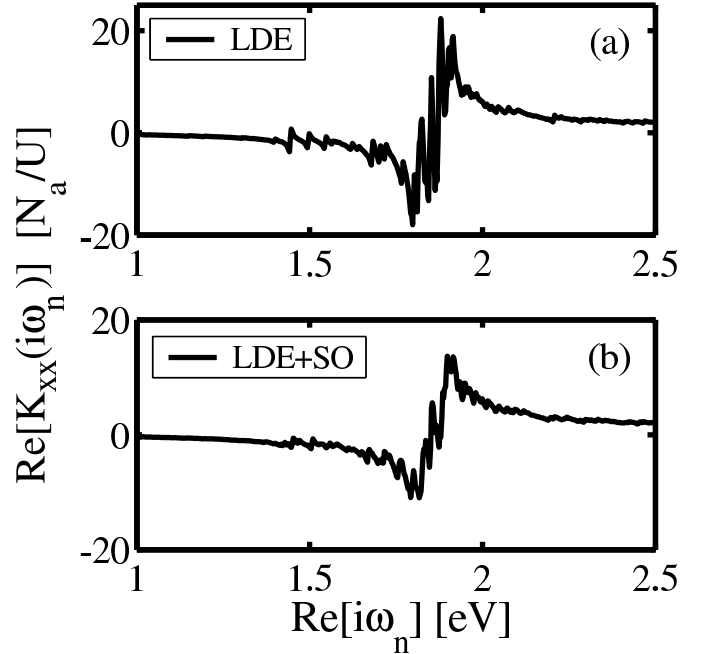}
 \caption{Real part of the transverse diagonal component of the kernel for
a hemispherical 26-atom nanoparticle, Local D-orbital Exchange model.
(a) Without SO interaction.
(b) With SO interaction.}
 \label{fig5_dsplitKernel}
 \end{figure}

 \begin{figure}
 \includegraphics[width=8.cm,height=8.cm]{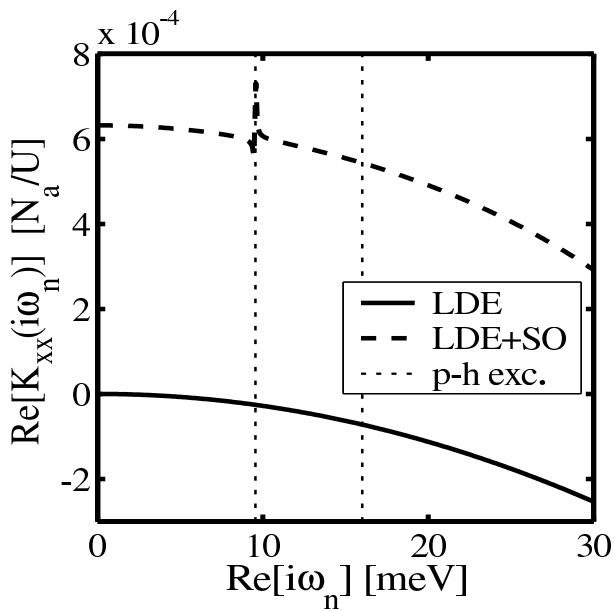}
\caption{Low-energy behavior of the Kernel for the same system as
in Fig.~\ref{fig5_dsplitKernel}. 
Solid line: without SO interaction; dashed line:
with SO interaction. The vertical dotted lines mark the position of 
particle-hole excitations.}
 \label{fig6_dsplitKernelzoom}
 \end{figure}

Similar results for the LDE model are shown in Fig.~(\ref{fig5_dsplitKernel}).
We can see [Fig.~\ref{fig5_dsplitKernel}(a)] that the kernel 
has already several poles even in absence of
SO coupling, since not all the energies of
majority- and minority-spin states are shifted by
the same rigid spin-splitting field, as they were in the case of the LRE model.
Most of the poles are still concentrated near $\Delta_{\rm MF}$, however.
The special properties of the LRE model which result in rigidly 
split mean-field bands do not, therefore, introduce and special 
artificial features in the low-energy excitation spectrum. 
From Fig.~\ref{fig5_dsplitKernel}(b) it would seem that
SO interaction do not introduce major effects, 
merely broadening the pre-existing pole structure.
In fact SO interaction changes the
analytical structure of the kernel at low energies, as it did for the LRE model.
 \begin{figure}
 \includegraphics[width=8.cm,height=8.cm]{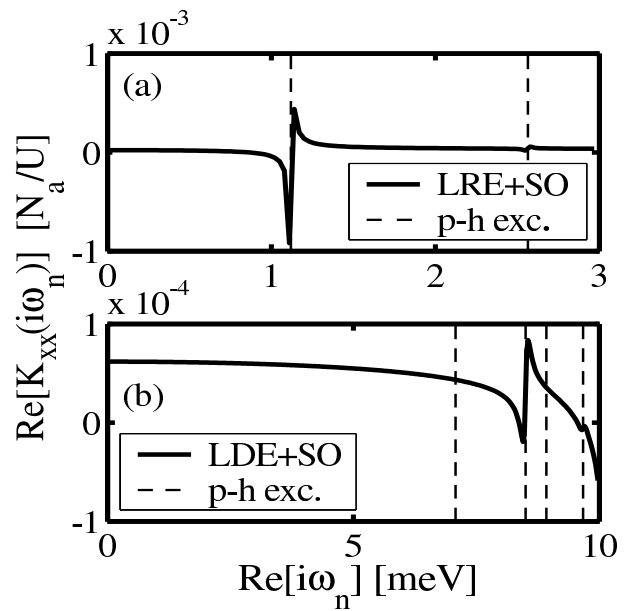}
\caption{Real part of the transverse diagonal component of the kernel at
low-energies, for
a hemispherical 143-atom nanoparticle. 
(a) Long-Range Exchange model.
(b) Local D-orbital Exchange model.
Vertical dashed lines mark the position of particle-hole excitations,
where ${\rm Re} K_{x\, x}$ has poles.}
 \label{fig7}
 \end{figure}
Indeed,  Fig.~\ref{fig6_dsplitKernelzoom} shows that when 
SO is included the first pole of $K_{x\, x}$
occurs again at an energy of the order of the 
single-particle mean level spacing. We find that most of these low energy
particle-hole excitations
involve pairs of states  that have {\it mostly} the same {\it minority} spin
character, as expected because essentially all states around
Fermi level are minority-spin states. Obviously matrix elements
$\langle J,\downarrow \vert s^{+} \vert I, \downarrow\rangle$ 
are identically zero and therefore there are no poles at these
low energies when SO coupling is not present.
It is only when SO coupling
is present that these states obtain a small admixture of majority-spin
character which produces non-zero matrix elements.
It is important to note that even when 
the rare presence of majority spin states near the Fermi level is recognized,
their matrix elements with minority spin states close in energy
are always very small 
because their orbital wavefunctions are almost orthogonal.
%
Fig.~\ref{fig7} shows the low-frequency behavior of the kernel for our two
models for the case of a 143-atom nanoparticle. We see essentially the
same trend as for the case of smaller nanoparticles: 
when SO is included the first pole of the Kernel
occurs at an energy of the order of the single-particle mean-level
spacing, which for this nanoparticle size is of the order of a few
meV's. Note that also at this particle size
the residue of this low-energy pole is three orders of magnitude smaller
than the residues of the poles occurring near $\Delta_{\rm MF}$.

The existence of spin-flip particle-hole excitations at energies
of order $\delta$ has important implications: we will show below that
these particle-hole pairs can couple to the 
low-energy spin collective mode when,
by increasing the nanoparticle size and/or by applying an external field, 
their energy start to approach $E_{\rm res}$.

\subsection{Spectral function: ${\rm Im}\chi_{x\, x}(\omega)$ 
vs. $\omega$}
\label{sp_func}

 \begin{figure}
 \includegraphics[width=8.cm,height=10.cm]{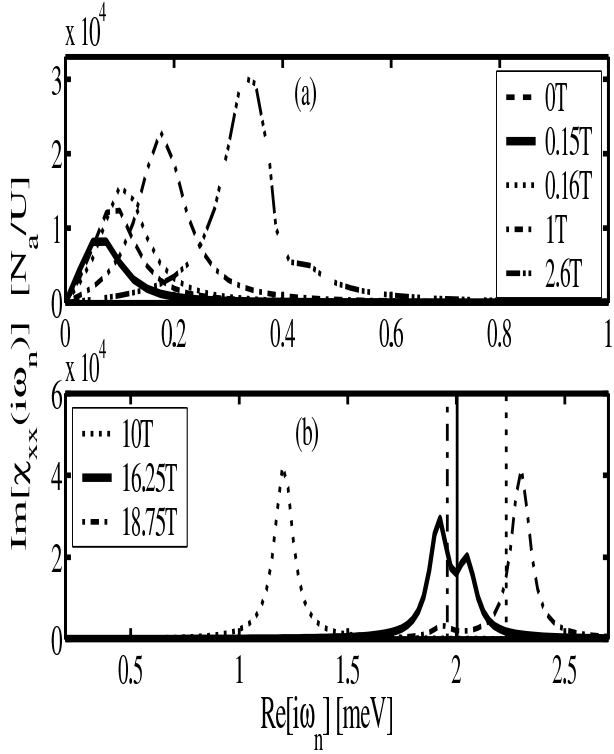}
 \caption{Imaginary part of the transverse diagonal spin susceptibility 
($x$ component) at different external magnetic fields,
for a hemispherical 143-atom nanoparticle in the LDE model.  
$\vec{H}_{\rm ext}$ is in the $ZX$-plane at an angle
$\pi/4$ with the $Z$-axis. ${H}_{\rm ext}= 0.15\, {\rm T}$ 
is a reversal point. 
This low-energy peak in the susceptibility
corresponds to a spin-collective mode
due to coherent magnetization fluctuations.
Vertical lines in (b) mark the position of the first
particle-hole excitation energy, which shifts down with increasing field.
When the collective mode energy approaches the particle-hole excitation energy,
the collective mode peak splits and part of its spectral weight is taken
away from the pole.}
 \label{fig8}
 \end{figure}

 \begin{figure}
\includegraphics{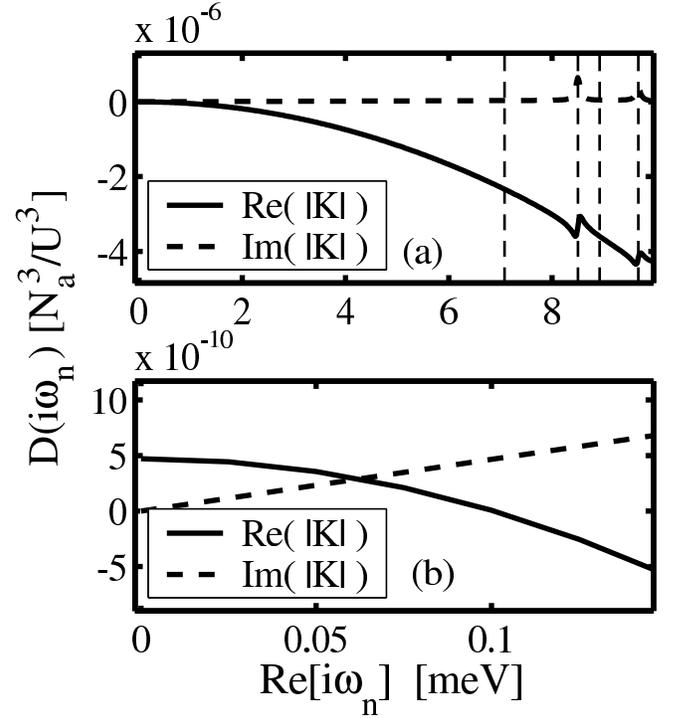}
 \caption{Determinant of the kernel defined in Eq.~\ref{det0}, for the
system of Fig.~\ref{fig8} at zero external field. The first zero 
of ${\rm Re}\,D(\omega)$ at 0.06 meV
corresponds to the ferromagnetic
resonance seen in Fig.~\ref{fig8}. ${\rm Re}\, D(\omega)$ has other zero's
very close to its poles that are located at particle-hole
energies (marked by vertical dashed lines).  The zero's correspond
to spin-flip particle-hole excitations, but their spectral weight 
in Fig.~\ref{fig8}
is virtually zero unless they approach the ferromagnetic resonance mode.}
 \label{fig9}
 \end{figure}

We now discuss the spin-fluctuation spectral function, that is the
imaginary part of the spin-susceptibility defined in Eq.~(\ref{susceptibility}).
In Fig.~\ref{fig8} we plot ${\rm Im}\,\chi_{x\,x}(\omega)$ as a function
of $\omega$ for a 143-atom hemispherical nanoparticle for several values
of the external magnetic field. As expected form our discussion in
Sec.~\ref{sec_gauss}, for this nanoparticle size we find that in general
the spectral function has only one pole that has substantial weight. 
At low external fields this pole occurs at a frequency that satisfies
Eq.~(\ref{det0}), well below the lowest particle-hole excitation. 
It can therefore be identified  as a spin-collective mode.
In this regime the collective mode is an exact elementary excitation,
at least within the Gaussian approximation that we consider.
The finite width in Fig.~\ref{fig8} is due to a finite $\eta$ in
$i\omega_n \to \omega + i\eta$. 

Apart from the ferromagnetic resonance
pole, the susceptibility has other poles at higher energies,
as seen in Fig.~\ref{fig9}, where we plot $D(\omega)$
given in Eq.~\ref{det0}.
These poles are all very close to MF particle-hole excitation
energies and have virtually zero spectral weight for this particle size,
when the external field is zero. We can therefore identify them
as the spin-flip particle-hole excitations discussed in Sec.~\ref{spectrum}.
 \begin{figure}
\includegraphics[width=8.cm,height=7.cm]{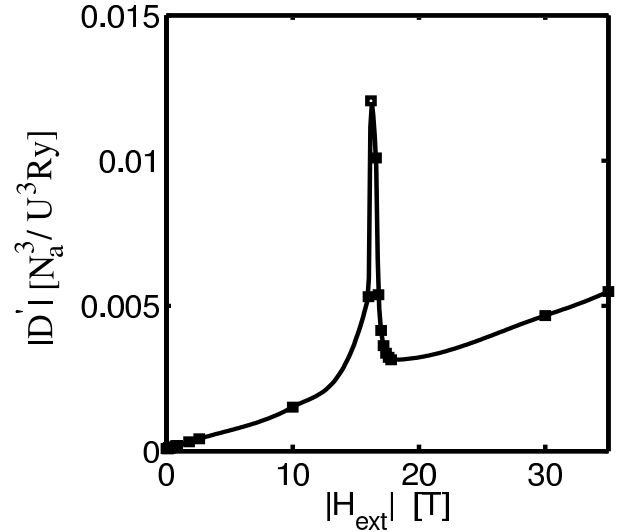}
 \caption{Derivative of the determinant of the kernel given in Eq.~\ref{det0},
calculated at $E_{\rm res}$, as a function of the external field
for the system of Fig.\ref{fig8}.
$D'(E_{\rm res})$ is inversely proportional to the residue of the
collective mode. The peak in $D'(E_{\rm res})$
occurs where the collective mode energy
and the first particle-hole excitation energy cross.}
 \label{der_det}
 \end{figure}

 \begin{figure}
 \includegraphics[width=8.cm,height=10.cm]{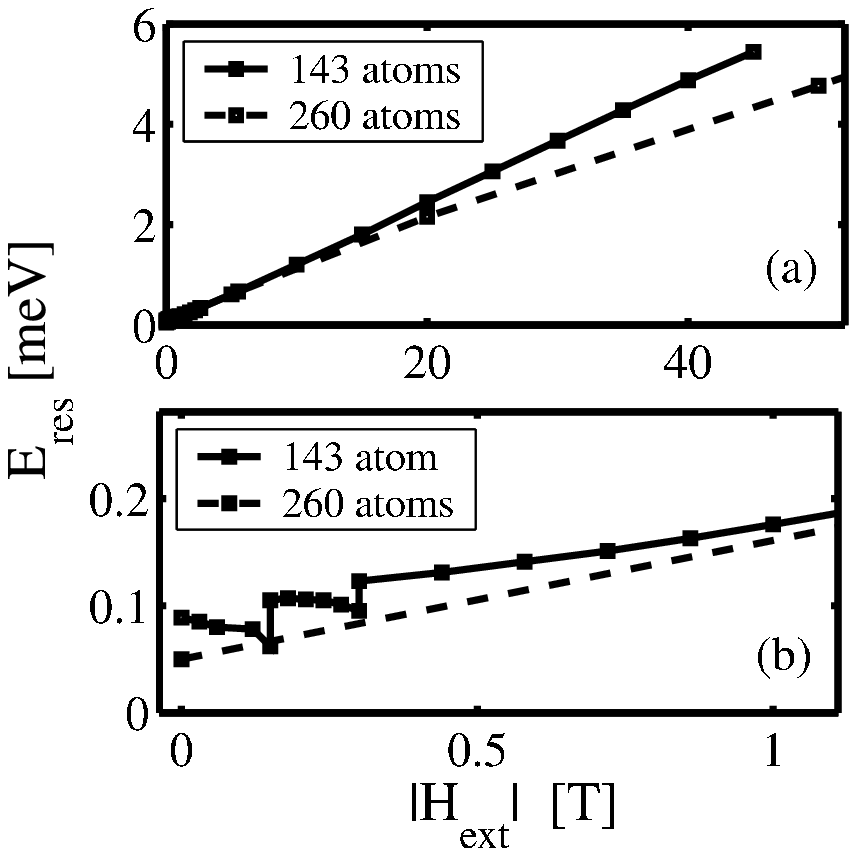}
 \caption{Variation of the collective-mode energy (low-energy peak 
in the susceptibility) as a function of the external field
for 143-atom and 260-atom nanoparticles.
$\vec{H}_{\rm ext}$ is in the $ZX$-plane at an angle $\pi/4$ with the 
$Z$-axis.
In (b) the detailed low-field behavior is shown.
The discontinuities for the 143-atom nanoparticle correspond to 
magnetization reversal points. (There are no reversal points in the
260-atom system for this external field direction.)}
 \label{figw_res}
 \end{figure}

As shown in Fig.~\ref{fig8},
we can use an external magnetic field to
manipulate both energy and spectral weight of the collective mode.
From zero field to the reversal field, the collective mode energy
and its spectral weight decrease. Beyond the coercive field they both start to
increase monotonically. As shown in Fig.~\ref{fig8}(b), if we keep increasing
the field, the collective mode energy starts to approach
the first particle-hole excitation energy, whose energy decreases
with increasing field. When the two energies are proximate to each other,
the collective-mode peak splits into two close resonances:
part of spectral weight of
the magnetization fluctuation is transferred to the 
nearby particle-hole excitation, as also illustrated in Fig.\ref{der_det}
where we plot the derivative of $D(\omega)$ at $E_{\rm res}$
as a function of the external field. The peak
in $D'$ corresponds to a sudden decrease of ${\cal Z}_{\rm res}$ 
given in Eq.~\ref{res_finite_w}:  
the collective mode can decay into a particle-hole excitation. 
More precisely, we can say that the two types of excitations are coupled,
and therefore it is no longer
meaningful to speak about spin-collective modes as distinct 
from spin-flip particle-hole excitations.
In our model the coupling mechanism is provided 
by the spin-orbit interaction. Indeed, we have checked that
without spin-orbit interaction, nothing happens to the collective peak
when its energy crosses a particle-hole excitation.

The coupling between collective modes and  particle-hole excitations
described above
mimics what would happen if, 
instead of manipulating the excitation energies with an external magnetic field,
we could progressively increase the size of the nanoparticle up to
a few thousand atoms.
In this case $E_{\rm res}$ would stay approximately constant
but $\delta$ would decrease. When these two energy scales cross
the lowest particle-hole
excitations will start to interact with the collective mode.
It is therefore expected that in a nanoparticle with a few thousand atoms
where $E_{\rm res} \approx \delta$, spin collective modes and particle-hole
pairs are strongly coupled.

Fig.~\ref{figw_res} shows the external field dependence of the energy of the
collective mode peak for a 143-atom and a 260-atom nanoparticle respectively.
The jumps in the energy for the 143-atom nanoparticle correspond to
reversal of the magnetic moment.
Interestingly, we have found that the peak energy follows accurately
Eq.~\ref{wres} for all values of the external field, even when 
its value is larger than the lowest particle-hole 
energy\footnote{The only exception is when a particle-hole excitation
energy exactly crosses the collective mode peak. 
As we have explained, in this case the two excitations
are coupled and there is a small splitting separating them.}.
The energy gap $E_{\rm res}$, obtained extrapolating the curve at
zero field, is approximately volume independent as expected from 
our general theory. We would like to point out that at 
low fields ($H_{\rm eff}\le 1 $T) 
$E_{\rm res} \sim 0.1$ meV, which of the order of the anisotropy constant/atom
$K$ in Cobalt\cite{stearns1986}, and is also
of the order of the tunneling resonance spacing observed 
in Refs.\onlinecite{gueron1999,deshmukh2001}.

\subsection{Anisotropic Fluctuations}

 \begin{figure}
 \includegraphics[width=8.0cm,height=6.cm]{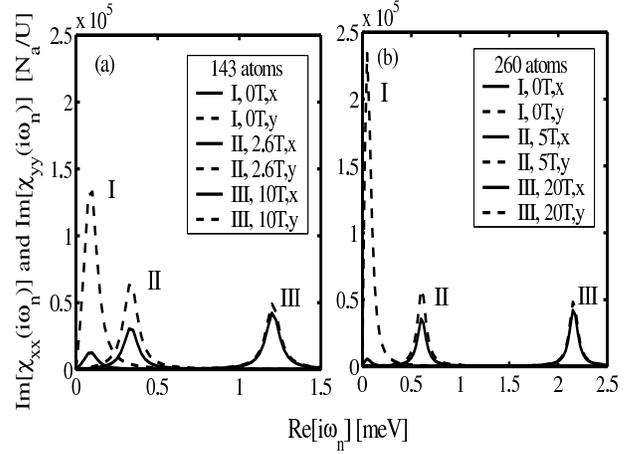}
 \caption{Imaginary part of the transverse diagonal spin susceptibility,
$x$ and $y$ components. The external magnetic field 
is in the $ZX$-plane at an angle
$\pi/4$ with the $Z$-axis.
At high external fields the two transverse components of the
spin susceptibility become equal. (a) Hemispherical 143-atom nanoparticle.
(b) Hemispherical 260-atom nanoparticle.}
 \label{fig12}
 \end{figure}

The dynamical susceptibility displays spatial anisotropy in its
transverse diagonal components due to the strong anisotropy of the
micromagnetic energy functional.
We illustrate this point in Fig.~\ref{fig12} by
plotting ${\rm Im}\,\chi_{x\,x}(\omega)$ together with
${\rm Im}\,\chi_{y\,y}(\omega)$.
We can see that also ${\rm Im}\,\chi_{y\,y}(\omega)$ has 
only one dominating collective-mode peak at the same
energy of the collective mode in 
${\rm Im}\,\chi_{x\,x}(\omega)$. This is obvious since in both
cases the collective mode energy is given by Eq.~(\ref{det0}).
However the {\it weight} of the pole in the two spectral functions
is different and its variation as a function of the external field
opposite. 
At large fields the difference between $x$ and $y$ components disappears
and the transverse susceptibility becomes isotropic. We find the same trend
for hemispherical nanoparticles containing 143 or 260 atoms. 


We can get an intuitive understanding of this behavior
by looking again at the classical micromagnetic energy functional
$E_{\rm tot}(\hat{\Omega})$ as a function 
of $\Theta$ and $\Phi$, shown in Fig.~ref{anisotropy}.
There are four equivalent minima in the $XY$ plane
at $\pm \hat X\pm \hat Y$, that is $\Theta=\pi/2$, $\Phi= \pi/4 +n\pi/2$,
$n=1,2,3$.
Note however that the energy barrier separating the four minima at
$\Theta=\pi/2$ is very low. We can interpret the collective mode as
the zero point motion of a two-dimensional anisotropic
harmonic oscillator, whose
potential is obtained by expanding  $E_{\rm tot}(\Theta,\Phi)$ to
second-order in $\Theta$ and $\Phi$ around one of the minima. 
Looking at the $x$-component of the spectral function
corresponds to exciting the collective oscillation mainly 
in the hard direction. We can pursue further this analogy and clarify
the effect of the external magnetic field. A magnetic field in the
$ZX$ plane starts to decrease the low barriers separating the four minima.
Thus the collective mode energy decreases. This continue up to
reversal.  
Note that at reversal the collective mode
energy does not go to zero because the potential landscape has only
a saddle point there: the potential still increases in the direction of the
two poles of the unit sphere
($\Theta = 0, \pi/2$).
After the last reversal has taken place, namely once the magnetization 
has flipped to a stable minimum,
a further increase of the magnetic field makes the spring constant
of the harmonic oscillator stiffer. Hence the collective mode
energy starts to increase. This behavior is summarized
in Fig.\ref{figw_res}.
At strong magnetic fields the
energy functional is dominated by the Zeeman term and any anisotropy
in the transverse fluctuations disappears.

Spherical particles have a more symmetric anisotropy energy landscape, 
without the strong XY easy plane anisotropy characteristic
of hemispherical particles. 
A reversal point in 
this case will correspond the disappearance of the energy barrier
in $E_{\rm tot}(\hat{\Omega})$
in all directions perpendicular to $\hat{\Delta}_{MF}$, or at least a 
less pronounced saddle point. Therefore spherical nanoparticles show 
a much stronger decrease of collective mode energies at reversal points.

\subsection{Berry Curvature}
\begin{figure}
 \includegraphics[width=9.cm,height=11.cm]{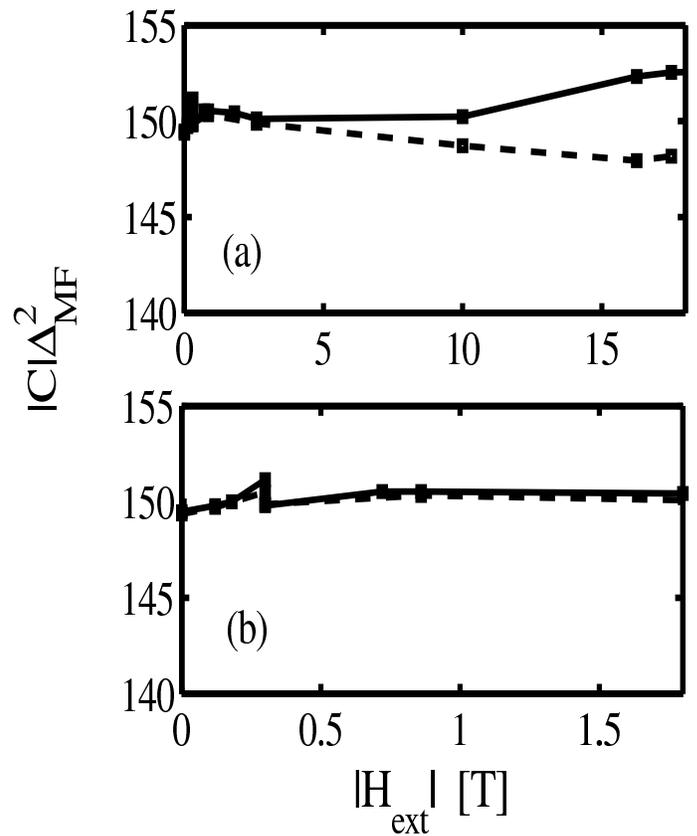}
 \caption{Berry curvature ${\cal C}(\Delta_{\rm MF})$ for 
a 143-atom nanoparticle, computed using two different methods:
the solid line is obtained from Eq.~\ref{berry_curvPT}; the dashed line
from Eqs.~\ref{triangle}-\ref{triangle2}. (b) Behavior at low fields.}
 \label{berryC_mf}
 \end{figure}

We conclude this session with a discussion of the numerical results
for the Berry curvature ${\cal C}(\Delta_{\rm MF})$ 
defined in Eq.~\ref{berrycurvature}. We have seen that 
${\cal C}(\Delta_{\rm MF})$ affects the quantization condition
of the collective mode energy given in Eq.~\ref{wres},
and is inversely proportional to its residue, Eq~\ref{resZ}.
Variations of ${\cal C}(\Delta_{\rm MF})$ from the constant
value $i {S/ \Delta_{\rm MF}^2}$ reflect the non-trivial 
role played by the spin-orbit interaction.
In Fig.~\ref{berryC_mf} we plot ${\cal C}(\Delta_{\rm MF})$ as
a function of an external magnetic field for a 143-atom nanoparticle.
We have computed ${\cal C}(\Delta_{\rm MF})$ in two different
ways, according to Eq.~\ref{berry_curvPT} 
and Eqs.~\ref{triangle}-\ref{triangle2} 
respectively.
The agreement between the two calculation methods is excellent,
especially at low fields [see fig.~\ref{berryC_mf}(b) ].
The fact that the computation method based on Eq.~\ref{triangle2} 
works well
is significant 
because this method
relies only on the knowledge of the ground-state for a given
magnetization orientation, which in principle can be obtained with fairly
good accuracy from density-functional calculations.
Thus Eq.~\ref{triangle2} provides a very convenient expression for computing
${\cal C}(\Delta_{\rm MF})$ beyond the mean-field approximation
used in the present work, and also for more realistic nanoparticle models.

From Fig.~\ref{berryC_mf} we can see that the external field dependence
of ${\cal C}(\Delta_{\rm MF})$ is rather smooth and weak, 
except when the system approaches a reversal point. At a reversal point
${\cal C}(\Delta_{\rm MF})$ suffers a discontinuous jump. 
On the other hand, ${\cal C}(\Delta_{\rm MF})$ is completely insensitive
to the crossings mentioned above between 
the ferromagnetic resonance mode and 
particle-hole excitations,
when the latter are of order $\approx \delta$.
In fact, if we look at the perturbative
expression of ${\cal C}(\Delta_{\rm MF})$ given in Eq.~\ref{berry_curvPT}, 
we can see that there is no reason to
expect any large fluctuations of this quantity when the smallest
energy denominator in the sum is $\approx \delta$.
If the collective mode energy approaches a particle-hole excitation,
${\cal C}(\Delta_{\rm MF})$ will still be well-defined and smooth,
but it will not tell us anything about how the collective mode spectral weight
is depleted in favor of the nearby particle-hole excitation.
In other words,  
the residue of the collective mode 
is inversely proportional to ${\cal C}(\Delta_{\rm MF})$
only when
the low-frequency expansion is valid, in which case it 
will be a smooth quantity.
When the collective mode energy is close to a particle-hole excitation energy of
order $\delta$, 
the low-frequency expansion is meaningless. 
One must look at the derivative of the determinant $D(\omega)$ calculated 
at the pole frequency, given in Eq.~\ref{res_finite_w}, to find the residue.
An important exception occurs when a particle-hole excitation energy
approaches zero 
(i.e. close to mean-field quasiparticle level crossings)\cite{ccmPRL03}.
These are events that can occur only for some particular orientations
of the magnetization and at isolated values of the external fields.
When the system is close to a mean-field quasiparticle level crossing, 
the Berry curvature landscape ${\cal C}(\Delta)$ is strongly distorted.
In this case the fluctuations of ${\cal C}(\Delta)$ are a direct indication
of the coupling between the collective mode and particle-hole excitations,
which obliterate the distinction between them.

\section{Conclusions}
\label{final}
In summary, we have developed a theory
of elementary spin excitations in ferromagnetic metal nanoparticles 
that provides a consistent and unified 
quantum description of both quasi-particle
and collective mode physics. Our formalism, based on a path integral
approach, allows us to make a connection between microscopic exchange
and spin-orbit interactions and classical micromagnetic theory.
We have shown that small nanoparticles have a 
collective excitation at energies below the lowest particle-hole
excitation energy, whose energy gap $E_{\rm res}$ is given
by the ratio of the anisotropy energy and the total Berry phase of the system. 
As the single-particle mean-level spacing decreases 
and becomes much smaller than $\sqrt{\alpha} E_{\rm res}$, the collective excited states evolves 
into a damped collective mode whose spectral weight is distributed over a 
large number of particle-hole excitations. 

We have illustrated these ideas by performing numerical calculations
of nanoparticles containing up to 260 atoms, described by a microscopic
tight-binding model. We have found that for this particle size 
there is typically an isolated collective excited state 
below the lowest particle-hole excitation which nearly exhausts the
spectral weight of the dynamical susceptibility.
The energy gap $E_{\rm res}$ is of the order of $0.1$meV, 
and is approximately independent of
the particle size.  Occasional crossings between 
primarily single-particle and collective excitations occur as a function of 
applied magnetic.  Near the crossing point, the collective mode peak splits,
as a result of resonant coupling between the two types of excitations
that is non-zero because of spin-orbit interactions.  These crossings become
more common as the particles become larger and cannot be avoided for 
system for nanoparticles with more than typically 10,000 atoms. 

Although a detailed comparison with the tunneling transport experiments
of Ref.~\onlinecite{gueron1999,deshmukh2001} is beyond the scope of 
the present paper, our analysis sheds light on  
some essential features which are quite relevant to the understanding
of the experimental results. What emerges from our theory is a
picture of elementary excitations that is far more complex
than that derived from earlier phenomenological models, 
where the effect of spin-orbit interaction is accounted for 
only indirectly by means of 
an uniaxial anisotropy term in a giant spin Hamiltonian that represents
the coherent magnetization dynamics.  In particular, our results suggest that
for the particle size considered in the experiments
quasi-particle and spin collective modes are most likely strongly
entangled by spin-orbit interactions. 
It is not unconceivable that such an intertwined set of excitations
could provide a rich low-energy tunneling spectrum, even in conditions
of equilibrium tunneling. 

The energy gap of the ferromagnetic resonance mode
can be viewed as the characteristic mean--energy-level spacing
between coupled collective-quasiparticle excitations described 
by an effective Hamiltonian. Interestingly enough, the value of
$E_{\rm res}$ deduced from our model calculations is
of the order of the observed tunneling resonance level spacing.
This value, which is also of the order of $K_{\rm bulk}$, is five
times larger than the anisotropy constant estimated from the measured
switching field using $K_{\rm sw}=\mu_B\,H_{\rm sw}$. 
The smallness of $K_{\rm sw}$ is one of the properties that led
the authors of 
Ref.\onlinecite{kleff_vdelft2001prb} to conclude that equilibrium spin
excitations involving only the lowest spin-multiplet cannot be resolved
in the present experiments and therefore cannot explain
the observed large density of resonances.
While the discrepancy between 
$K_{\rm sw}$ and $K_{\rm bulk}$ is 
still a puzzle\cite{mandar_thesis2002} (see  however \footnote{
One possible explanation that we have put forward in 
Ref.~\cite{ac_cmc_ahm2002} is that under the action of an external field,
small groups of atoms
can change their orientation relative to other parts of the nanoparticle.
The small $K_{\rm sw}$ extracted from tunneling experiments would then
refer to such local reorientation processes.}),
our results suggest that the these low-lying excitations
are in fact being detected.
These considerations do not necessarily imply that the main features
of the tunneling experiments can be understood by equilibrium
transitions alone. Measurements on gated devices\cite{deshmukh2001} 
convincingly support
the hypothesis that non equilibrium transitions play a crucial role.
The point that we want to make here is that
low-lying {\it equilibrium spin excitations}
are probably equally important for the interpretation of the
experimental results.

It would be highly desirable to perform new tunneling experiments in more
controlled situations and especially with smaller nanoparticles,  so that
the spin excitations associated with 
the electronic degrees of freedom and with the magnetization collective
coordinate could be more easily disentangled.  The physical condition required
to reach this regime is that the single-particle mean level spacing be
larger than the total anisotropy energy\cite{ccmPRL03}.
For Cobalt this would
imply dealing with nanoparticles containing of the order of 100 atoms,
which is within reach of the present experimental capabilities.

\section{Acknowledgements}
We would like to thank Mandar Deshmukh and Dan Ralph for several
discussions about their experimental results,
Jan von Delft and Piet Brouwer for stimulating and informative interactions.
This work was supported in part by the Swedish Research Council
under Grant No:621-2001-2357, by the faculty of natural science 
of Kalmar University,
and in part by the National Science Foundation
under Grants DMR 0115947 and DMR 0210383. 
Support from the Office of Naval Research under Grant N00014-02-1-0813
is also gratefully acknowledged.

\bibliography{../biblio/Biblio}

\end{document}